%% file: main.tex
\documentclass[11pt,a4paper]{article}
\usepackage[T1]{fontenc}
\usepackage[utf8]{inputenc}
\usepackage{lmodern,amsmath,amssymb,graphicx,booktabs,array}
\usepackage[left=23mm,right=23mm,top=22mm,bottom=24mm]{geometry}
\usepackage[authoryear,round]{natbib}
\usepackage{xcolor}
\definecolor{linkink}{RGB}{36,74,95}
\usepackage[colorlinks=true,allcolors=linkink,bookmarksnumbered=true]{hyperref}
\hypersetup{pdftitle={Solving hyperbolic conservation laws and learning solution operators using integral-potential dual residuals and gradient downweighting},pdfauthor={Fansheng Xiong and Jiawei Liu},pdfsubject={Hyperbolic conservation laws and solution-operator learning}}
\begin{document}
\begin{center}
{\LARGE\bfseries Solving hyperbolic conservation laws and learning solution operators using integral-potential dual residuals and gradient downweighting\par}
\vspace{9pt}
{\large Fansheng Xiong$^{1}$ \qquad Jiawei Liu$^{2}$\par}
\vspace{5pt}
{\small $^{1}$Beijing Institute of Mathematical Sciences and Applications,\\
Beijing 101408, China\par}
\vspace{3pt}
{\small $^{2}$Advanced Institute for Materials Research, Tohoku University,\\
Sendai, Japan\par}
\end{center}
\input{abstract}
\input{introduction}

\input{methods}

\input{operator}

\input{single_results}
\input{operator_results}
\input{discussion}
\input{references}
\input{appendix_a}
\input{appendix_b}
\end{document}

%% file: abstract.tex
\begin{abstract}

In neural approximations of hyperbolic conservation laws, resolving a thin transition layer does not necessarily locate the discontinuity accurately. Gradient downweighting facilitates the formation of sharp fronts, but a downweighted strong-form residual alone can leave their positions inaccurate. We develop a gradient-weighted physics-informed neural network (gwPINN) based on an integral potential and extend it to learning solution operators for parameterized initial-value problems. The conserved state is recovered by spatial differentiation of the potential. Training combines a potential residual, which constrains transport through interval conservation, with its spatial derivative, the strong-form residual. The two residuals are downweighted separately. We test the method on the one-dimensional Burgers equation, Euler Riemann problems, and the two-dimensional Burgers equation. In two paired moving-shock ablations, omitting the potential residual increases the terminal position error by factors of about 83 and 14; in the first pair, the layer width changes by less than 1\%. For operator learning, a shared network conditioned on initial data is trained with both residuals and a small set of terminal profiles. With 25 labeled initial conditions, the method gives lower terminal errors than our RiemannONet implementation on all 100 low-pressure-ratio Euler test inputs for each of three seeds, and on 99, 100, and 100 pressure-function inputs, respectively. For the latter family, the same architecture trained without PDE residuals fits the terminal profiles but gives larger errors at unsupervised intermediate times. The results demonstrate the value of combining the two residuals for front localization and time-dependent prediction in these experiments. Local negative pressures at higher pressure ratios remain unresolved.

\end{abstract}

%% file: introduction.tex
\section{Introduction}\label{sec:intro}

Hyperbolic conservation laws govern a wide range of transport, wave propagation, and compressible-flow problems. Their solutions can develop discontinuities in finite time, even for smooth initial data. The speed of a shock is governed by the Rankine--Hugoniot relation, while the relevant entropy condition selects the physically admissible solution \citep{r01,r02}. Numerical approximations must therefore resolve discontinuities, predict their positions accurately, and capture their interactions.

Finite-volume methods enforce discrete conservation by sharing numerical fluxes between adjacent cells \citep{r03}. Weighted essentially non-oscillatory schemes adapt their reconstruction stencils to retain accuracy in smooth regions and suppress oscillations near discontinuities \citep{r04}. Entropy-viscosity methods vary the numerical dissipation according to the local entropy residual \citep{r05}. Euler shock-tube problems, which typically contain a shock, a contact discontinuity, and a rarefaction wave, are standard tests of these methods \citep{r06}. For the Euler equations, preserving positive density and pressure is also essential \citep{r07}.

Neural networks offer an alternative representation of PDE solutions as continuous functions. Early work embedded the initial and boundary conditions in a neural trial function and determined its parameters by minimizing the equation residual \citep{r08}. The deep Galerkin method instead uses randomly sampled collocation points in training \citep{r09}. In physics-informed neural networks (PINNs), equation residuals, initial and boundary conditions, and available observations enter a common objective \citep{r10}. The resulting network can be evaluated at arbitrary coordinates, and the governing equation provides constraints where solution data are unavailable. This property is useful for forward and inverse problems and for data fusion \citep{r11,r12}. Automatic differentiation and software such as DeepXDE facilitate the implementation \citep{r13}.

A particular difficulty arises when the solution contains a discontinuity. A smooth neural approximation replaces a shock with a transition layer; as that layer narrows, the large derivatives can dominate the strong-form loss and hinder training. This difficulty has been observed in PINNs for high-speed flows \citep{r14}. Gradient imbalance, unequal convergence rates of error components, and ill-conditioning of the loss compound the problem \citep{r15,r16,r17,r18}, motivating improvements in network parameterization, optimization, and collocation sampling \citep{r19,r20}.

Gradient-weighted PINNs (gwPINNs) address part of this difficulty by reducing the weight of the strong-form residual in compression regions, allowing thinner shock layers to develop \citep{r21,r22}. However, downweighting the residual near a shock can also weaken the constraint on its position. For a jump of fixed amplitude, a displaced shock and a broadened shock can both give large field errors, although the underlying errors are quite different. It is thus useful to measure the shock-position error and transition-layer width separately, alongside the usual field-error norms.

Conservation and weak formulations provide additional ways to constrain the solution. Conservative PINNs match fluxes and solutions at subdomain interfaces \citep{r23}. Space--time control-volume methods incorporate conservation, entropy, and total variation into the loss \citep{r24}. The hp-VPINN combines variational residuals with domain decomposition \citep{r25}, and wPINNs employ Kru\v{z}kov entropies and test functions for scalar conservation laws \citep{r26}. Other approaches construct weak residuals using boundary and entropy fluxes \citep{r27}, or combine finite-volume balances with downweighted strong-form residuals \citep{r28}. These studies motivate the use of integral conservation information together with local differential constraints.

Of particular relevance to the present work are integral PINNs, in which a neural network approximates a spatial primitive of the conserved state. Rajvanshi and Ketcheson \citep{r29} used such a representation to improve the prediction of shock positions and speeds. Coupled integral PINNs train separate networks for the potential and the state \citep{r30}. We instead obtain the state directly by differentiating one potential and impose both the potential equation and the strong form of the original conservation law. For a differentiable representation, the strong-form residual is the spatial derivative of the potential residual, a construction related to gradient-enhanced PINNs \citep{r31}. We downweight the two residuals independently to retain the thin shock layers attainable with gwPINNs while constraining cumulative transport.

A PINN trained for one set of initial data normally has to be retrained when the initial condition changes. For applications requiring repeated solves, it is natural to learn the mapping from initial data to solution functions instead. Operator learning trains a shared set of network parameters over a family of input conditions and allows subsequent evaluation for previously unseen inputs. Operator approximation theory \citep{r32}, neural reduced-order models for parameterized PDEs \citep{r33}, and neural operators \citep{r34} provide the foundations for such an approach. The computational advantage is expected mainly in repeated evaluations after training.

DeepONet encodes the input function and the query coordinates in separate branch and trunk networks \citep{r35}. The Fourier neural operator (FNO) uses integral layers implemented in Fourier space \citep{r36}. The approximation properties of these architectures have been analyzed in \citep{r37,r38}, and subsequent developments include factorized Fourier neural operators and convolutional neural operators \citep{r39,r40}. Physics-informed DeepONet and PINO incorporate governing equations into operator training, so the solution can also be constrained at space--time points without labels \citep{r41,r42}.

In operator learning for conservation laws, the locations of discontinuities vary with the input. Representing such families in fixed linear bases can require many modes. Nonlinear reduced-order models \citep{r43}, the nonlinear decoder in NOMAD \citep{r44}, and nonlinear reconstruction for PDEs with discontinuities \citep{r45} address this issue. For Euler Riemann problems, RiemannONets map pressure parameters to terminal solution profiles through two-stage training \citep{r46}. EGPT-PINN employs pretraining, coordinate transformations, and entropy regularization for parameterized nonlinear conservation laws \citep{r47}. Conservation structures have also been encoded directly in neural operators \citep{r48}.

Large datasets such as PDEBench \citep{r49} and pretraining approaches such as Poseidon \citep{r50} have broadened the scope of PDE operator learning. Here we consider a more limited data setting: only a subset of the initial conditions has a reference solution, and these references are given at the terminal time. The aim is nevertheless to predict solutions throughout the time interval. A good terminal fit does not by itself establish accuracy at intermediate times. Comparisons must also account for the grid used to measure errors near discontinuities and the implementation of the baseline \citep{r51}.

Our approach begins with an integral-potential dual-residual gwPINN for a single initial condition. We then use the same potential representation and physical losses to train a conditional network over a family of initial conditions. In the operator model, the initial-data encoding is included in the input, collocation points are sampled jointly over the initial-data and space--time domains, and a data loss is evaluated on the available terminal profiles. One trained network can then be used for different initial conditions.

The experiments focus on two questions: whether the potential residual improves discontinuity localization and whether the two residuals improve time-dependent predictions for parameterized problems. Sections~2 and 3 describe the single-problem solver and its extension to operator learning. Section~4 presents one- and two-dimensional Burgers tests and Euler Riemann problems; Section~5 considers scalar-pressure and pressure-function inputs. Section~6 discusses the results and limitations.

%% file: methods.tex
\section{Integral-potential PINN for a single initial-value problem}\label{sec:pinn}

\subsection{Integral potential and conservation}\label{sec:potential}

Consider a system of hyperbolic conservation laws in $d$ spatial dimensions,

\begin{equation}\label{eq:1}
\partial_t U+\sum_{j=1}^{d}\partial_{x_j}F_j(U)=0,\qquad U(0,x)=U_0(x),\qquad (t,x)\in(0,T)\times\Omega.
\end{equation}

Here $U\in\mathbb{R}^m$ is the conserved state and $F_j$ is the flux in direction $j$. Hyperbolicity means that, for a physically admissible state and any unit vector $n$, the flux Jacobian $\sum_j n_jF_j'(U)$ is diagonalizable with real eigenvalues. The initial data are denoted by $U_0$, and boundary conditions are specified for each problem. We begin with the one-dimensional case.

Let $\Omega=(a,b)$ and $F=F_1$. For a bounded weak solution with boundary flux traces, define the integral potential

\begin{equation}\label{eq:2}
\Phi(t,x)=\int_a^xU(t,\xi)\,d\xi-\int_0^tF(U(s,a^+))\,ds.
\end{equation}

The notation $a^+$ refers to the interior trace at the left endpoint: $U(s,a^+)=\lim_{x\downarrow a}U(s,x)$ when the one-sided limit exists. Likewise, $b^-$ refers to the interior trace at the right endpoint. Conservation over $(a,x)$ gives $\partial_t\int_a^x U\,d\xi=F(U(t,a^+))-F(U(t,x))$. It follows from Eq.~(2) that $\Phi_x=U$ and $\Phi_t=-F(U)$ almost everywhere. The boundary-flux integral fixes the time-dependent integration constant of the potential.

The potential remains continuous across a jump in $U$. For a piecewise smooth solution, differentiation of its two traces along the shock trajectory $x=X(t)$ yields $[\Phi_t]+\dot X[\Phi_x]=0$, giving the Rankine--Hugoniot relation $\dot X[U]=[F(U)]$. Here $[\cdot]$ denotes the right value minus the left value. If the endpoint states remain constant, the boundary potentials are $\Phi(t,a)=-tF(U_L)$ and $\Phi(t,b)=\int_a^bU_0(\xi)\,d\xi-tF(U_R)$.

The integral representation is sensitive to a displacement of a front. Consider an isolated jump with fixed left and right states. Moving it by $\delta x$ changes the integral across the front by $(U_L-U_R)\delta x$. In contrast, a symmetric broadening of the front can leave that integral unchanged. This difference motivates the use of both an integral-potential residual and a local strong-form residual.

\subsection{Network representation and the two residuals}\label{sec:dual}

In a conventional PINN, the network approximates the state $U_\theta(t,x)$ directly. The strong-form residual $\partial_tU_\theta+\partial_xF(U_\theta)$ is evaluated at collocation points without solution labels, and the network parameters $\theta$ are optimized together with the initial and boundary losses \citep{r10,r13}. In our formulation, a fully connected network $N_\theta$ approximates a correction to a prescribed potential extension $B$ that incorporates the initial data:

\begin{equation}\label{eq:3}
\Phi_\theta(t,x)=B(t,x)+tN_\theta(t,x),\qquad U_\theta(t,x)=\partial_x\Phi_\theta(t,x).
\end{equation}

For a single initial jump at $x_0$, we use

\begin{equation}\label{eq:4}
\begin{aligned}
B(t,x)&=\int_a^x\left\{U_L+(U_R-U_L)S\left(\frac{\xi-x_0}{wt}\right)\right\}\,d\xi,\\
S(z)&=\frac12(1+\tanh z),\qquad t>0.
\end{aligned}
\end{equation}

The coefficient $w$ is set to 0.6 times the largest absolute characteristic speed on either side of the initial jump. The width of the transition in $B$ grows linearly with time; its subsequent shape and propagation are adjusted by the network correction. For multiple jumps, their contributions are superposed, and smooth initial data are incorporated through their spatial primitive. Provided that $\partial_xN_\theta$ stays bounded as $t\to0$, Eq.~(3) satisfies the initial condition at continuity points of $U_0$.

Differentiating Eq.~(3) with respect to space gives the state. We define two residuals,

\begin{equation}\label{eq:5}
R_\Phi=\partial_t\Phi_\theta+F(U_\theta),
\end{equation}

\begin{equation}\label{eq:6}
R_s=\partial_t U_\theta+\partial_xF(U_\theta)=\partial_xR_\Phi.
\end{equation}

Evaluating $R_\Phi$ requires first derivatives of the potential; $R_s$ requires second derivatives. Both residuals are computed from the same network. On any fixed interval $[\ell,r]\subset[a,b]$, the following identity holds:

\begin{equation}\label{eq:7}
\frac{d}{dt}\int_\ell^rU_\theta\,dx+F(U_\theta(t,r))-F(U_\theta(t,\ell))=R_\Phi(t,r)-R_\Phi(t,\ell).
\end{equation}

Equation~(7) expresses the interval conservation defect as the difference between the potential residuals at the two endpoints. Equation~(6) identifies the strong-form residual as their local spatial derivative. Penalizing both is useful: a large, slowly varying $R_\Phi$ may have a small derivative, whereas even a small $R_\Phi$ can have a large derivative if it varies rapidly. The boundary values $\Phi(t,a)=-\int_0^tF(U(s,a^+))\,ds$ and $\Phi(t,b)=\int_a^bU_0(\xi)\,d\xi-\int_0^tF(U(s,b^-))\,ds$ prescribe the potential at the endpoints and fix its spatially constant part. The identities above hold for a differentiable potential representation. The interpolation used at time-slab restarts introduces an approximation.

\subsection{Gradient downweighting and spatial sampling}\label{sec:weight}

Let $(t_i,x_i)$, $i=1,\ldots,N_f$, denote the interior collocation points, $D=\operatorname{diag}(d_1,\ldots,d_m)$ a fixed positive component scaling, and $G_i=G_\theta(t_i,x_i)\geq0$ a compression indicator. The quadrature weights are $a_i=1$ for uniform sampling; under importance sampling, $a_i$ is the reciprocal of the relative sampling density. The physical losses are

\begin{equation}\label{eq:8}
\begin{aligned}
\mathcal L_\Phi&=\frac1{N_f}\sum_{i=1}^{N_f}a_i\sum_{k=1}^{m}
\left(\frac{R_{\Phi,k}(t_i,x_i)}{d_k(1+c_\Phi G_i)}\right)^2,\\
\mathcal L_s&=\frac1{N_f}\sum_{i=1}^{N_f}a_i\omega_i\sum_{k=1}^{m}
\left(\frac{R_{s,k}(t_i,x_i)}{d_k(1+c_s G_i)}\right)^2.
\end{aligned}
\end{equation}

\begin{equation}\label{eq:9}
\mathcal L_{\mathrm{PINN}}=\mathcal L_\Phi+\lambda_s\mathcal L_s+\lambda_b\mathcal L_b.
\end{equation}

For $\mathcal L_b$, we sum the sample means of $\|D^{-1}(\Phi_\theta-\Phi^{\mathrm{bc}})\|_2^2$ at the two endpoints. No separate initial-condition loss is needed in the main model because Eq.~(3) incorporates the initial data. The factor $\omega_i$ in Eq.~(8) weights the strong-form residual in time. It equals 1 in the operator experiments; some single-problem experiments use an early-time window. To see why, consider a fan of width $O(t)$ whose leading strong-residual term is $A((x-x_0)/t)/t$. Integrating its square in space gives a term of order $1/t$. A weight $\omega(t)\propto t^p$ with $p>0$ makes this contribution integrable in time.

Following \citet{r21}, gradient downweighting is applied to the residual before squaring it. We use distinct coefficients $c_\Phi$ and $c_s$ for the two residuals. For the one-dimensional Burgers equation, $G_\theta=(-\partial_xU_\theta)_+$, where $(z)_+=\max(z,0)$. For the Euler equations, we use $G_\theta=\{(-\partial_xv_\theta)_+|\partial_xp_\theta|+10^{-12}\}^{1/2}$, with velocity $v$ and pressure $p$. Since velocity and pressure are constant across a pure contact discontinuity, this indicator downweights the residuals less strongly there. Density protection is used in evaluating the Euler flux and its derivatives.

Adaptive sampling and gradient downweighting serve different purposes. The former places more collocation points in regions of large steepness or residual \citep{r20}, with quadrature weights $a_i$ correcting for the nonuniform sampling density. The latter changes the residual penalty itself in regions of compression. We choose the two independently. The network and its loss terms are illustrated in Fig.~1.

\begin{figure}[!htbp]\centering
\input{figure01.tex}
\caption{Integral-potential dual-residual network. The single-problem model takes space--time coordinates $(t,x)$ as input, while operator learning adds the initial-data encoding $e$. The network output and the prescribed extension form the potential, whose derivatives give the state and the two residuals. The connections are schematic; the actual depths and widths are specified for each experiment.}\label{fig:fig01}
\end{figure}
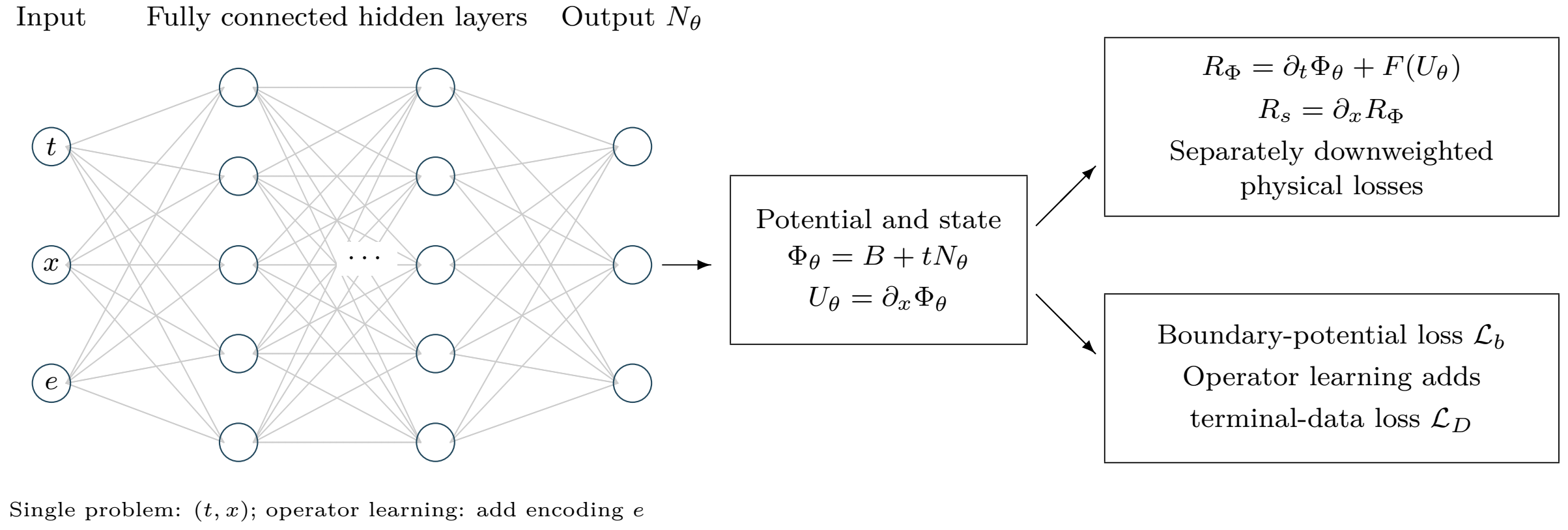

\subsection{Directional potentials in multiple dimensions}\label{sec:multi}

For multiple spatial dimensions, choose initial directional potentials $P_j^0\in\mathbb{R}^m$ such that $\sum_j\partial_{x_j}P_j^0=U_0$. Define $P_j(t,x)=P_j^0(x)-\int_0^tF_j(U(s,x))\,ds$. Taking the divergence and using Eq.~(1) gives $\sum_j\partial_{x_j}P_j=U$. With $P_{j,\theta}=B_j+tN_{j,\theta}$, the state and residuals become

\begin{equation}\label{eq:10}
U_\theta=\sum_j\partial_{x_j}P_{j,\theta},\qquad R_j=\partial_tP_{j,\theta}+F_j(U_\theta),\qquad R_s=\sum_j\partial_{x_j}R_j.
\end{equation}

In the corresponding potential loss, the squared one-dimensional residual (after component scaling) in Eq.~(8) is replaced by $\sum_j\|D^{-1}R_j\|_2^2$. The initial directional potentials are fixed for each problem, and their boundary values follow from time integrals of the known boundary flux traces. For the two-dimensional Burgers tests in this work, the traces are obtained from one-dimensional boundary problems.

%% file: figure01.tex
\setlength{\unitlength}{1mm}%
\begin{picture}(164,55)%
\color{black}\linethickness{0.12mm}%
{\color{gray!40}\qbezier(11.5,14)(19.5,11.0)(27.5,8)}%
{\color{gray!40}\qbezier(11.5,14)(19.5,15.5)(27.5,17)}%
{\color{gray!40}\qbezier(11.5,14)(19.5,20.0)(27.5,26)}%
{\color{gray!40}\qbezier(11.5,14)(19.5,24.5)(27.5,35)}%
{\color{gray!40}\qbezier(11.5,14)(19.5,29.0)(27.5,44)}%
{\color{gray!40}\qbezier(11.5,26)(19.5,17.0)(27.5,8)}%
{\color{gray!40}\qbezier(11.5,26)(19.5,21.5)(27.5,17)}%
{\color{gray!40}\qbezier(11.5,26)(19.5,26.0)(27.5,26)}%
{\color{gray!40}\qbezier(11.5,26)(19.5,30.5)(27.5,35)}%
{\color{gray!40}\qbezier(11.5,26)(19.5,35.0)(27.5,44)}%
{\color{gray!40}\qbezier(11.5,38)(19.5,23.0)(27.5,8)}%
{\color{gray!40}\qbezier(11.5,38)(19.5,27.5)(27.5,17)}%
{\color{gray!40}\qbezier(11.5,38)(19.5,32.0)(27.5,26)}%
{\color{gray!40}\qbezier(11.5,38)(19.5,36.5)(27.5,35)}%
{\color{gray!40}\qbezier(11.5,38)(19.5,41.0)(27.5,44)}%
{\color{gray!40}\qbezier(30.5,8)(39.0,8.0)(47.5,8)}%
{\color{gray!40}\qbezier(30.5,8)(39.0,12.5)(47.5,17)}%
{\color{gray!40}\qbezier(30.5,8)(39.0,17.0)(47.5,26)}%
{\color{gray!40}\qbezier(30.5,8)(39.0,21.5)(47.5,35)}%
{\color{gray!40}\qbezier(30.5,8)(39.0,26.0)(47.5,44)}%
{\color{gray!40}\qbezier(30.5,17)(39.0,12.5)(47.5,8)}%
{\color{gray!40}\qbezier(30.5,17)(39.0,17.0)(47.5,17)}%
{\color{gray!40}\qbezier(30.5,17)(39.0,21.5)(47.5,26)}%
{\color{gray!40}\qbezier(30.5,17)(39.0,26.0)(47.5,35)}%
{\color{gray!40}\qbezier(30.5,17)(39.0,30.5)(47.5,44)}%
{\color{gray!40}\qbezier(30.5,26)(39.0,17.0)(47.5,8)}%
{\color{gray!40}\qbezier(30.5,26)(39.0,21.5)(47.5,17)}%
{\color{gray!40}\qbezier(30.5,26)(39.0,26.0)(47.5,26)}%
{\color{gray!40}\qbezier(30.5,26)(39.0,30.5)(47.5,35)}%
{\color{gray!40}\qbezier(30.5,26)(39.0,35.0)(47.5,44)}%
{\color{gray!40}\qbezier(30.5,35)(39.0,21.5)(47.5,8)}%
{\color{gray!40}\qbezier(30.5,35)(39.0,26.0)(47.5,17)}%
{\color{gray!40}\qbezier(30.5,35)(39.0,30.5)(47.5,26)}%
{\color{gray!40}\qbezier(30.5,35)(39.0,35.0)(47.5,35)}%
{\color{gray!40}\qbezier(30.5,35)(39.0,39.5)(47.5,44)}%
{\color{gray!40}\qbezier(30.5,44)(39.0,26.0)(47.5,8)}%
{\color{gray!40}\qbezier(30.5,44)(39.0,30.5)(47.5,17)}%
{\color{gray!40}\qbezier(30.5,44)(39.0,35.0)(47.5,26)}%
{\color{gray!40}\qbezier(30.5,44)(39.0,39.5)(47.5,35)}%
{\color{gray!40}\qbezier(30.5,44)(39.0,44.0)(47.5,44)}%
{\color{gray!40}\qbezier(50.5,8)(59.0,11.0)(67.5,14)}%
{\color{gray!40}\qbezier(50.5,8)(59.0,17.0)(67.5,26)}%
{\color{gray!40}\qbezier(50.5,8)(59.0,23.0)(67.5,38)}%
{\color{gray!40}\qbezier(50.5,17)(59.0,15.5)(67.5,14)}%
{\color{gray!40}\qbezier(50.5,17)(59.0,21.5)(67.5,26)}%
{\color{gray!40}\qbezier(50.5,17)(59.0,27.5)(67.5,38)}%
{\color{gray!40}\qbezier(50.5,26)(59.0,20.0)(67.5,14)}%
{\color{gray!40}\qbezier(50.5,26)(59.0,26.0)(67.5,26)}%
{\color{gray!40}\qbezier(50.5,26)(59.0,32.0)(67.5,38)}%
{\color{gray!40}\qbezier(50.5,35)(59.0,24.5)(67.5,14)}%
{\color{gray!40}\qbezier(50.5,35)(59.0,30.5)(67.5,26)}%
{\color{gray!40}\qbezier(50.5,35)(59.0,36.5)(67.5,38)}%
{\color{gray!40}\qbezier(50.5,44)(59.0,29.0)(67.5,14)}%
{\color{gray!40}\qbezier(50.5,44)(59.0,35.0)(67.5,26)}%
{\color{gray!40}\qbezier(50.5,44)(59.0,41.0)(67.5,38)}%
\put(10,14){\color{linkink}\circle{3.8}}%
\put(10,14){\makebox(0,0){\scriptsize$e$}}%
\put(10,26){\color{linkink}\circle{3.8}}%
\put(10,26){\makebox(0,0){\scriptsize$x$}}%
\put(10,38){\color{linkink}\circle{3.8}}%
\put(10,38){\makebox(0,0){\scriptsize$t$}}%
\put(29,8){\color{linkink}\circle{3.8}}%
\put(29,17){\color{linkink}\circle{3.8}}%
\put(29,26){\color{linkink}\circle{3.8}}%
\put(29,35){\color{linkink}\circle{3.8}}%
\put(29,44){\color{linkink}\circle{3.8}}%
\put(49,8){\color{linkink}\circle{3.8}}%
\put(49,17){\color{linkink}\circle{3.8}}%
\put(49,26){\color{linkink}\circle{3.8}}%
\put(49,35){\color{linkink}\circle{3.8}}%
\put(49,44){\color{linkink}\circle{3.8}}%
\put(69,14){\color{linkink}\circle{3.8}}%
\put(69,26){\color{linkink}\circle{3.8}}%
\put(69,38){\color{linkink}\circle{3.8}}%
\put(10,51){\makebox(0,0){\scriptsize Input}}%
\put(39,51){\makebox(0,0){\scriptsize Fully connected hidden layers}}%
\put(69,51){\makebox(0,0){\scriptsize Output $N_\theta$}}%
\put(39,26){\colorbox{white}{\scriptsize$\cdots$}}%
\put(38,1){\makebox(0,0){\tiny Single problem: $(t,x)$; operator learning: add encoding $e$}}%
\put(72,26){\vector(1,0){5}}%
\put(79,18){\framebox(30,17){\shortstack{\scriptsize Potential and state\\[2pt]\scriptsize$\Phi_\theta=B+tN_\theta$\\[2pt]\scriptsize$U_\theta=\partial_x\Phi_\theta$}}}%
\put(110,30){\vector(1,1){6}}%
\put(110,23){\vector(1,-1){6}}%
\put(117,31){\framebox(46,18){\shortstack{\scriptsize$R_\Phi=\partial_t\Phi_\theta+F(U_\theta)$\\[2pt]\scriptsize$R_s=\partial_xR_\Phi$\\[2pt]\scriptsize Separately downweighted\\\scriptsize physical losses}}}%
\put(117,6){\framebox(46,17){\shortstack{\scriptsize Boundary-potential loss $\mathcal L_b$\\[2pt]\scriptsize Operator learning adds\\[2pt]\scriptsize terminal-data loss $\mathcal L_D$}}}%
\end{picture}%

%% file: operator.tex
\section{Solution-operator learning for parameterized initial-value problems}\label{sec:operator}

We now consider a family of initial conditions $U_0(\cdot;\mu)$ rather than a single initial condition. The potential representation, two residuals, and gradient downweighting are unchanged, but the input to the network also includes an encoding of the initial data. Collocation points are sampled over both the initial-data and space--time domains, and the same parameters are used for the entire input family. A small set of terminal profiles supplements the physical losses. Once trained, the conditional network predicts a new initial condition without further optimization.

\subsection{An integral potential conditioned on initial data}\label{sec:conditional}

Let $\mathcal S:\mu\mapsto U(\cdot,\cdot;\mu)$ denote the solution mapping. In our experiments the flux is fixed, and only the initial data vary. For a scalar input, the parameter itself serves as the encoding $e(\mu)$. A function input $h$ is represented by its values at fixed sensors, $e(h)=(h(z_1),\ldots,h(z_M))$. The conditional potential and solution mapping are then

\begin{equation}\label{eq:11}
\Phi_\theta(t,x;e)=B(t,x;e)+tN_\theta(t,x,e),\qquad \mathcal S_\theta(\mu)(t,x)=\partial_x\Phi_\theta(t,x;e).
\end{equation}

The network takes the concatenation of $(t,x)$ and $e$ as input, has six hidden layers of width 30 with $\tanh$ activations, and returns a correction for each of the three conserved components. There are three inputs for the scalar-pressure Euler model and 18 for the function-input model with 16 sensors. Thus we use a conditional fully connected network to represent the solution mapping \citep{r32,r34}, in contrast to the separate branch and trunk networks of DeepONet \citep{r35}.

For each initial condition, Eqs.~(5) and (6) are evaluated by differentiating with respect to space and time while holding $e$ fixed. To make a prediction, we construct the encoding and extension $B$ for the new initial data and evaluate the potential and its spatial derivative. Scalar inputs give a parameter-to-function map; function inputs give a map on the family of functions resolved by the sensors.

The extension $B$ also needs to account for the variation of function inputs between sensors. The variable initial pressure in our experiments is quadratic on a prescribed interval. We fit a cubic spline to the 16 pressure readings and incorporate its primitive in the initial energy potential. The network sees the sensor readings, although the input sampler draws the three coefficients of the quadratic function.

\subsection{Terminal supervision and physical losses}\label{sec:training}

Suppose that each of $n$ training initial conditions has a reference profile at the terminal time $T$, with $M_x$ spatial points per profile. Let $U_{ik}^{\mathrm{lab}}$ denote the reference conserved state and $e_i$ the corresponding initial-data encoding. The data and total losses are

\begin{equation}\label{eq:12}
\mathcal L_D=\frac1{nM_x}\sum_{i=1}^{n}\sum_{k=1}^{M_x}\left\|D^{-1}\left(\partial_x\Phi_\theta(T,x_k;e_i)-U_{ik}^{\mathrm{lab}}\right)\right\|_2^2,
\end{equation}

\begin{equation}\label{eq:13}
\mathcal L_{\mathrm{op}}=\mathcal L_\Phi+\lambda_s\mathcal L_s+\lambda_b\mathcal L_b+\lambda_D\mathcal L_D.
\end{equation}

The reference density, velocity, and pressure are converted to conserved variables before computing the data loss. For the physical losses in Eq.~(8), collocation points $(t_i,x_i,e_i)$ are drawn over initial conditions both with and without terminal labels. In contrast, the data loss is evaluated only for the initial conditions with known terminal profiles. Throughout the paper, the label count refers to the number of such initial conditions, not the number of individual spatial observations. This training strategy follows the principle of physics-informed operator learning \citep{r41,r42}.

The scalar- and function-input networks have 4863 and 5313 parameters, respectively. Each run uses 40000 interior Latin-hypercube collocation points and 400 boundary points at each endpoint. In the function-input case, sampling takes place jointly over $(t,x)$ and the three initial-data coefficients. During physical training, interior points are redrawn uniformly every 100 steps, and the temporal weight is 1. We first perform 1000 Adam warm-up steps with a learning rate of $10^{-3}$, taking $H$ and $t\partial_xH$ as targets for $N_\theta$ and $t\partial_xN_\theta$, respectively. The target $H=-(1-S)F_L(x)-SF_R$ is constructed from the initial fluxes, with the smooth transition $S$ in Eq.~(4) using a dimensionless width $w(t+0.002)$. We then minimize Eq.~(13) using double-precision, full-memory self-scaled BFGS, with an iteration budget of 3000. The loss coefficients are $c_\Phi=0.1$, $c_s=1$, $\lambda_s=0.1$, and $\lambda_b=\lambda_D=100$. Component scales are obtained from reference initial states and fluxes and are fixed within each input family.

We consider three label counts, $n=25,100,400$. Each of the two scalar-pressure families from the public RiemannONets dataset \citep{r46} has 400 training and 100 test inputs, with 200 spatial points per profile. For $n<400$, we sort the training parameters, divide them into $n$ equal-sized rank intervals, and select the parameter at the middle rank of each interval. We also evaluate the same test inputs against exact Riemann solutions on a 4097-point grid. For the pressure-function family, finite-volume solutions provide reference profiles for 400 training, 80 validation, and 100 test functions. All methods use the first $n$ training functions from the same pool. Training and test profiles contain 301 and 6145 spatial points, respectively.

In preliminary low-pressure-ratio experiments, we used 80 validation parameters outside the original parameter grid to choose the data-loss weight and numerical precision. The same settings are used for the other input families. To isolate the effect of the PDE residuals, we train a control network of the same architecture without $\mathcal L_\Phi$ or $\mathcal L_s$. This control retains the initial-data extension, initial-flux warm-up, boundary potentials, and terminal data; its BFGS objective is $\lambda_b\mathcal L_b+\lambda_D\mathcal L_D$.

We also compare with a local implementation of the two-stage DeepONet from RiemannONets \citep{r46}. The branch and trunk networks each have five hidden layers of width 150. Our Rowdy activation combines a scaled $\tanh$ with trainable sine functions, initialized with zero amplitudes and phases. The public implementation recommends cosine terms for the two scalar-pressure benchmarks, so our results refer specifically to the sine variant. The trunk network is trained for 12000 steps on normalized terminal primitive variables. From 150 candidate spatial modes, a basis is obtained by singular value decomposition with a relative truncation tolerance of $10^{-5}$. With the basis fixed, expansion coefficients for the training profiles are obtained by double-precision least squares. The branch network is then trained for 30000 steps; checkpoints are selected by validation error.

All three methods see the same labeled initial conditions and are tested on the same inputs. The RiemannONet comparison is a comparison of terminal accuracy at equal label counts, not at equal architecture size or optimization cost. To assess the contribution of the physical losses to predictions at intermediate times, we use the control with the same architecture.

%% file: single_results.tex
\section{Numerical experiments for single initial-value problems}\label{sec:single}

\subsection{Error metrics and statistical reporting}\label{sec:metrics}

For a scalar profile $v$ evaluated on a uniform grid, we use the relative errors $E_q(v)=\|v_\theta-v_{\mathrm{ref}}\|_q/\|v_{\mathrm{ref}}\|_q$, $q=1,2$. In the Euler tests, errors are computed separately for density, velocity, and pressure, then averaged arithmetically. We locate a front by its half-amplitude crossing and define its transition-layer width as the distance between the 10\% and 90\% crossings. These quantities are interpolated near the reference front and reported in the original spatial coordinate.

Independent training runs are treated as replicates; the tables report means and sample standard deviations. Table~1 includes the four seeds with lower terminal relative errors out of five, whereas Table~3 and the operator statistics include all corresponding seeds. Table~4 reports the two-dimensional results for seed~1, the same models plotted in Figs.~10 and S5. For the four-shock problem, seed~1 has the lowest terminal error among the three original runs.

The one-dimensional plots show pointwise averages of the runs listed in their captions. For each isolated-contact problem, seed~2, which has the largest density overshoot at the plotted times, is excluded from the average in the figure but included in Table~3. All error metrics are calculated from the raw predictions of the individual networks, not from the averaged profiles.

\subsection{One-dimensional Burgers equation}\label{sec:burgers}

We consider the Burgers equation $u_t+(u^2/2)_x=0$ on $[-1,1]$, extending the domain to $[-1,6]$ for the square pulse. Boundary states are prescribed for each case. We use six hidden layers of width 30, 10000 interior collocation points, 3000 Adam steps, and 2500 self-scaled BFGS steps. The compression indicator is $(-u_x)_+$, with potential and strong-form downweighting coefficients of 0.1 and 10, respectively. Table~1 summarizes the terminal errors.

\begin{table}[!htbp]\centering
\caption{Terminal relative errors (\%) for one-dimensional Burgers problems. Each group includes four seeds and reports the mean $\pm$ sample standard deviation.}\label{tab:1}
\small\setlength{\tabcolsep}{3pt}\renewcommand{\arraystretch}{1.18}
\begin{tabular*}{\linewidth}{@{\extracolsep{\fill}}lccc@{}}\toprule
Initial data or wave type & $T$ & \shortstack[l]{$E_1$\\Four seeds} & \shortstack[l]{$E_2$\\Four seeds}\\\midrule
Moving shock $(1,0)$ & 1 & $0.0229 \pm 0.00501$ & $0.621 \pm 0.0871$\\
Stationary shock $(1,-1)$ & 1 & $0.0364 \pm 0.00182$ & $1.21 \pm 0.0564$\\
\shortstack[l]{Rarefaction $(-1,1)$\\Original configuration} & 1 & $0.0411 \pm 0.00336$ & $0.0714 \pm 0.00394$\\
\shortstack[l]{Rarefaction $(-1,1)$\\Revised configuration} & 1 & $0.00497 \pm 0.000414$ & $0.00727 \pm 0.000957$\\
Shock merging $(1,0.5,0)$ & 1.4 & $0.0899 \pm 0.0174$ & $1.43 \pm 0.155$\\
Square pulse & 4 & $0.381 \pm 0.177$ & $2.45 \pm 0.214$\\
Sinusoidal wave & 1 & $0.604 \pm 0.0254$ & $6.78 \pm 0.165$\\
\bottomrule\end{tabular*}
\end{table}

The initial states of the moving shock are 1 and 0, giving a propagation speed of $1/2$. In all five runs, the position error at the sampled times does not exceed $8.8\times10^{-5}$, and the terminal layer width lies between $6.09\times10^{-4}$ and $1.02\times10^{-3}$. Figure~2 shows close agreement with the exact front $x=t/2$. With the right state changed to $-1$, the shock is stationary. The maximum position error in each run is below $9.1\times10^{-5}$ (Fig.~S1).

\begin{figure}[!htbp]\centering
\includegraphics[width=\linewidth]{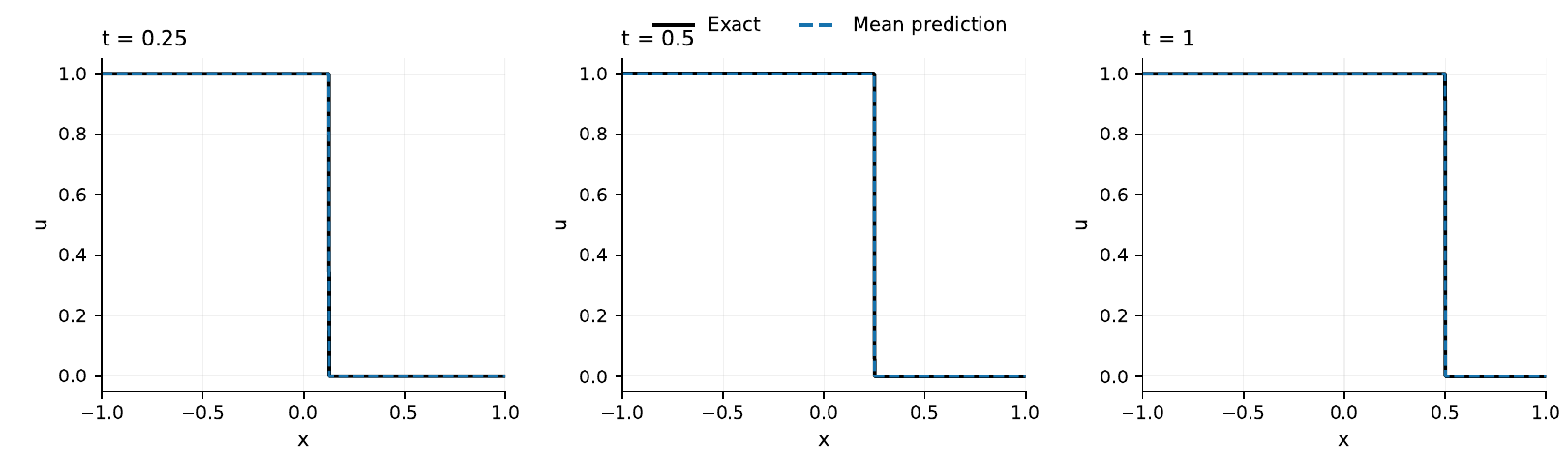}
\caption{Full-domain profiles of the moving Burgers shock at $t=0.25,0.5,1$. Black solid lines denote the exact solution, and blue dashed lines denote the mean predictions of seeds 1, 2, 3, and 5.}\label{fig:fig02}
\end{figure}

For the centered rarefaction, the intermediate state $u=x/t$ joins the two constant states at $x=\pm t$. With a linear early-time weight and mixed steepness--residual sampling (Fig.~3), terminal $E_2$ is below 0.01\% for all five seeds, compared with 0.0680\%--0.0838\% for the original configuration. Both the temporal weight and the sampling rule were changed in this comparison.

\begin{figure}[!htbp]\centering
\includegraphics[width=\linewidth]{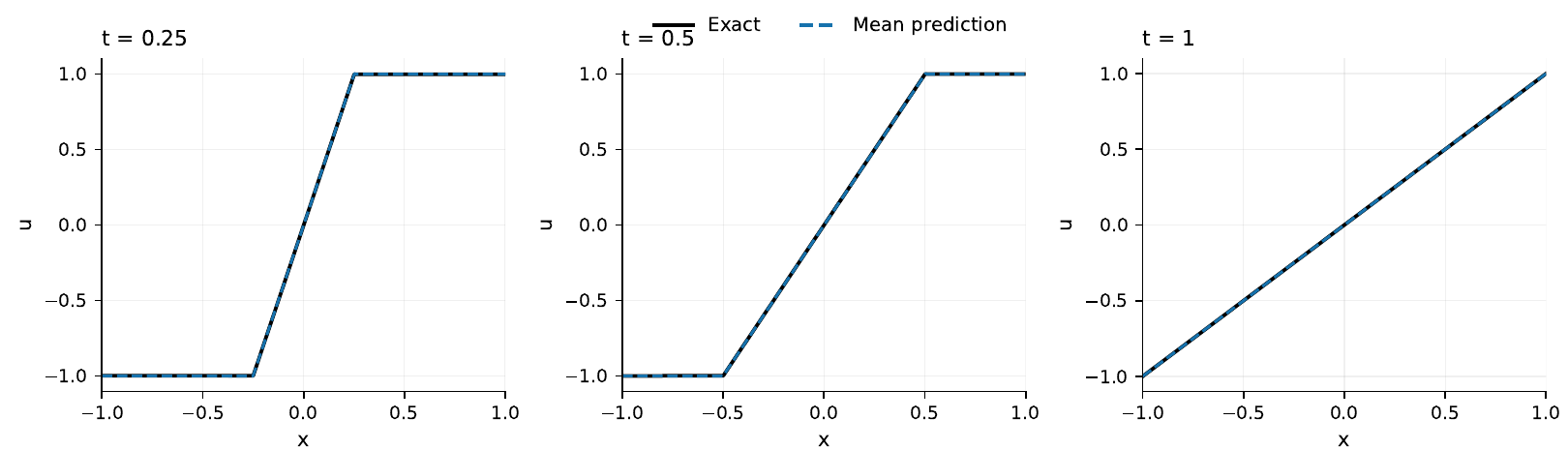}
\caption{Full-domain profiles of the centered rarefaction at $t=0.25,0.5,1$. Black solid lines denote the exact solution, and blue dashed lines denote the mean predictions of seeds 2, 3, 4, and 5 under the revised configuration.}\label{fig:fig03}
\end{figure}

In the shock-merging case, the initial jumps are located at $-0.5$ and 0 and propagate at speeds 0.75 and 0.25, respectively. They meet at $(x,t)=(0.25,1)$, and the merged shock reaches $x=0.45$ at $T=1.4$. Predicted profiles are shown in Fig.~S2.

For the square pulse $u_0=\mathbf1_{[0,1]}$, the rarefaction overtakes the shock at $t=2$; thereafter $x_s(t)=\sqrt{2t}$. With time-slab restarts, maximum position errors over the five seeds range from 0.00058 to 0.00150, compared with 0.0061--0.0069 for whole-interval training (Fig.~4). At each restart, the predicted potential and state are transferred from the previous slab. Mature fronts are locally reconstructed using a hyperbolic tangent to preserve their resolution. Seed~1 uses a revised criterion for accepting this reconstruction, while seeds 2--5 use the earlier criterion. Each subinterval requires additional optimization.

For the smooth initial condition $u_0=-\sin(\pi x)$, the algorithm estimates the shock-formation time from the reciprocal compressive slope at $t=0.25$. The estimates from five runs lie between 0.3167 and 0.3190, compared with the exact formation time $1/\pi\approx0.3183$. At the terminal time, the layer widths are 0.0082--0.0122, and field errors exceed those for a single initial jump (Fig.~S3).

\begin{figure}[!htbp]\centering
\includegraphics[width=\linewidth]{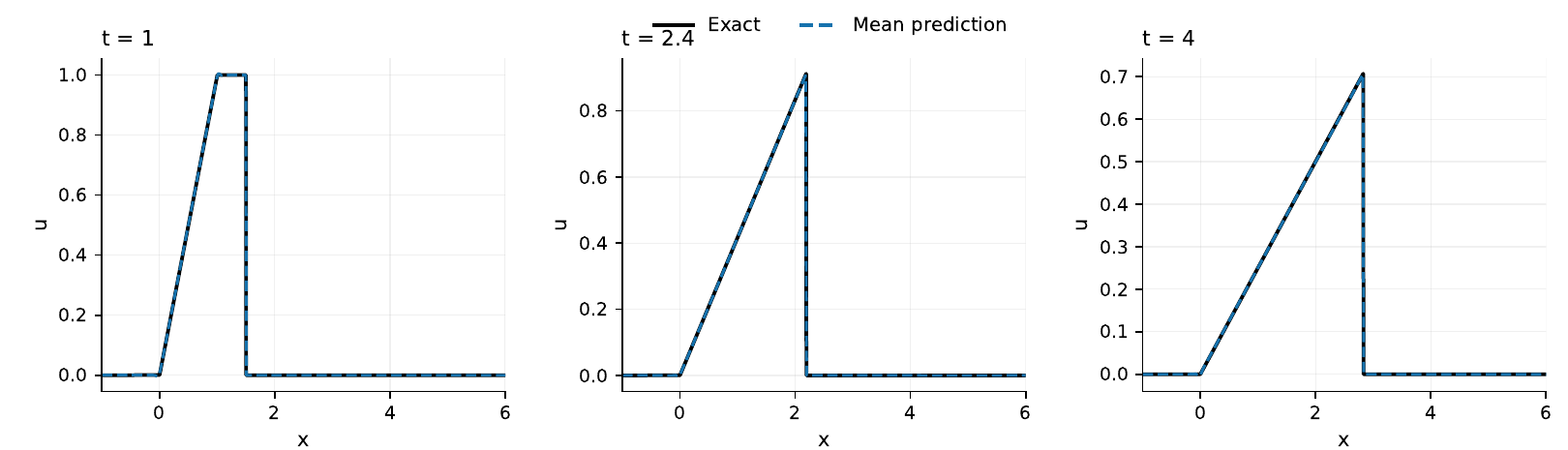}
\caption{Square-pulse profiles at $t=1,2.4,4$. Black solid lines denote the exact solution, and blue dashed lines denote the mean predictions of the four time-slab models with lower terminal errors.}\label{fig:fig04}
\end{figure}

\subsection{Paired ablations of the potential residual for shock localization}\label{sec:paired}

We examine the contribution of the potential residual in two paired moving-shock experiments. Within each pair, we keep the initial and boundary information, potential representation, downweighted strong-form residual, and temporal weight unchanged, omitting only the potential residual. In the first pair, the terminal layer width changes by less than 1\%, but the position error increases from $7.8\times10^{-5}$ to $6.49\times10^{-3}$ (Table~2). In the second pair, the error increases from $1.76\times10^{-4}$ to $2.49\times10^{-3}$. The first pair is particularly informative: two fronts of almost identical width can differ substantially in position accuracy. The results support the use of the potential residual for localization. Additional Burgers ablations of residual downweighting, initial-condition treatment, temporal weights, and sampling appear in Fig.~S4.

\begin{table}[!htbp]\centering
\caption{Moving-shock ablation experiments, including both seed pairs. Terminal absolute position errors and 10\%--90\% layer widths are measured at 32000 equally spaced cell centers, in the same units as $x$.}\label{tab:2}
\small\setlength{\tabcolsep}{3pt}\renewcommand{\arraystretch}{1.18}
\begin{tabular*}{\linewidth}{@{\extracolsep{\fill}}lcccc@{}}\toprule
&\multicolumn{2}{c}{Terminal position error}&\multicolumn{2}{c}{Terminal width}\\
\cmidrule(lr){2-3}\cmidrule(lr){4-5}
Seed&With $R_\Phi$&Without $R_\Phi$&With $R_\Phi$&Without $R_\Phi$\\\midrule
1 & 0.000078 & 0.00649 & 0.00240 & 0.00242\\
2 & 0.000176 & 0.00249 & 0.00188 & 0.00242\\
\bottomrule\end{tabular*}
\end{table}

In a separate comparison, four paired runs use either a fixed refinement band or dynamic refinement. Both members of each pair start from identical weights, share 5000 uniform collocation points, and use another 5000 refined points. The fixed band is obtained from the starting prediction over $0.15T\leq t\leq T$; dynamic refinement updates sampling probabilities every 100 steps using the current steepness and residual. Each run receives 500 Adam steps and 2000 self-scaled BFGS steps. Dynamic refinement narrows the terminal layer in all four pairs, with a median reduction of 18.0\%. The position error decreases in three pairs, with a median reduction of 45.5\%; in the fourth, it increases from $2.75\times10^{-5}$ to $6.23\times10^{-5}$. On 48 independent space--time control volumes at three scales, the normalized root-mean-square balance defect after dynamic refinement is 0.350--0.428 times its initial value. Nevertheless, checks at five Kru\v{z}kov entropy levels reveal local positive entropy defects. The training constraints do not guarantee the entropy inequality.

\subsection{Euler Riemann problems and isolated contact discontinuities}\label{sec:euler}

On $[-0.5,0.5]$, we use the ideal-gas Euler flux,

\[U = (\rho,\rho v,E)^{\mathsf{T}},\quad\quad F(U) = (\rho v,\rho v^{2} + p,v(E + p))^{\mathsf{T}},\]

\[p = (\gamma - 1)\left( E - \frac{1}{2}\rho v^{2} \right),\quad\gamma = 1.4.\]

The initial interface is at the origin. The left and right states of the Sod problem are $(\rho,v,p)_L=(1,0,1)$ and $(\rho,v,p)_R=(0.125,0,0.1)$; exchanging them gives the reverse Sod problem. The moving-gas case uses $(0.445,0.698,0.7)$ and $(0.5,0,0.571)$, producing two shocks. Increasing the left pressure to 3.528 gives the Lax problem. The downweighting indicator is the velocity--pressure gradient combination in Section~2.3, with $c_\Phi=0.1$ and $c_s=1$.

\begin{table}[!htbp]\centering
\caption{Statistics over all seeds for the Euler problems, reported as mean $\pm$ standard deviation. Field errors are arithmetic means of terminal relative errors in density, velocity, and pressure. For the final column, each seed's maximum position error is first taken over identifiable shocks and contacts at sampled times $t\geq T/4$, and then summarized across seeds.}\label{tab:3}
\small\setlength{\tabcolsep}{3pt}\renewcommand{\arraystretch}{1.18}
\begin{tabular*}{\linewidth}{@{\extracolsep{\fill}}lccccc@{}}\toprule
Initial data & $T$ & Seeds & $E_1$ (\%) & $E_2$ (\%) & \shortstack[l]{Max. position error\\($\times10^{-3}$)}\\\midrule
Sod & 0.20 & 10 & $0.218 \pm 0.0615$ & $1.27 \pm 0.453$ & $0.507 \pm 0.182$\\
Reverse Sod & 0.20 & 10 & $0.248 \pm 0.0857$ & $1.71 \pm 0.689$ & $0.664 \pm 0.342$\\
Moving gas & 0.20 & 10 & $0.22 \pm 0.027$ & $1.38 \pm 0.166$ & $2.35 \pm 0.659$\\
Lax & 0.13 & 10 & $0.551 \pm 0.135$ & $2.59 \pm 0.335$ & $0.249 \pm 0.178$\\
\shortstack[l]{Contact\\(denser left)} & 0.20 & 5 & $0.0168 \pm 0.00191$ & $0.314 \pm 0.0266$ & $0.00674 \pm 0.00661$\\
\shortstack[l]{Contact\\(denser right)} & 0.20 & 5 & $0.0242 \pm 0.00312$ & $0.442 \pm 0.0412$ & $0.0363 \pm 0.0281$\\
\bottomrule\end{tabular*}
\end{table}

For each Euler problem, we use 10000 interior collocation points, 1000 warm-up steps based on the smoothed initial flux, and 3000 self-scaled BFGS steps. Figures~5--8 show the predicted shocks, contacts, and rarefactions at half-time and at the terminal time. In Sod's problem, the maximum position error at the sampled times $t\geq T/4$ is below $9.2\times10^{-4}$ in all ten runs. The terminal pressure-shock widths are 0.000245--0.00113, whereas the density-contact widths are 0.00973--0.0183. Contacts are also broader in the reverse Sod and Lax problems. The moving-gas problem has the largest mean maximum position error in Table~3, about 0.00235.

\begin{figure}[!htbp]\centering
\includegraphics[width=\linewidth]{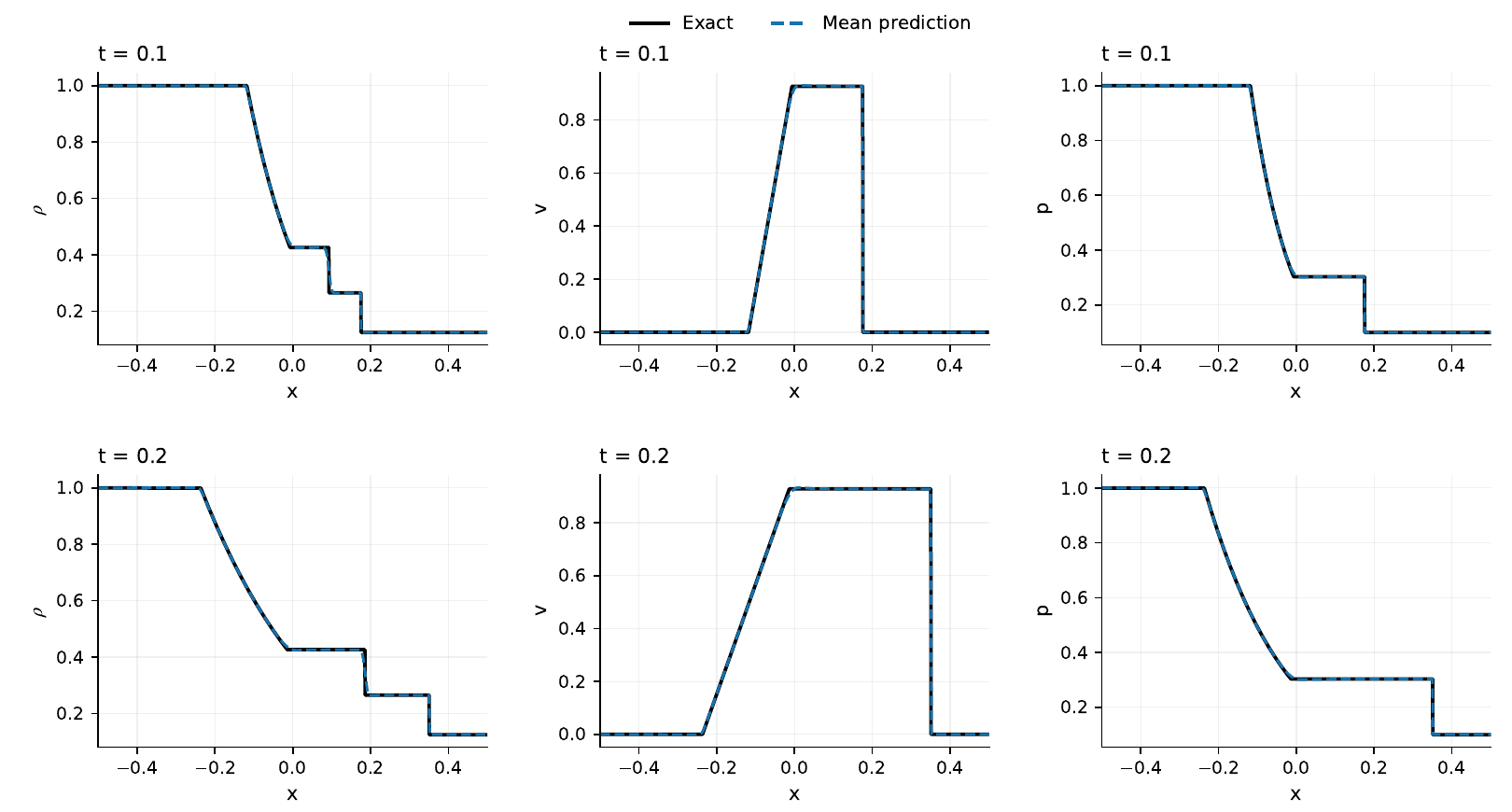}
\caption{Sod shock tube. The top and bottom rows correspond to $T/2$ and $T=0.2$; the columns show density, velocity, and pressure. Black solid lines denote the exact solution, and blue dashed lines denote the mean predictions over all ten seeds.}\label{fig:fig05}
\end{figure}

\begin{figure}[!htbp]\centering
\includegraphics[width=\linewidth]{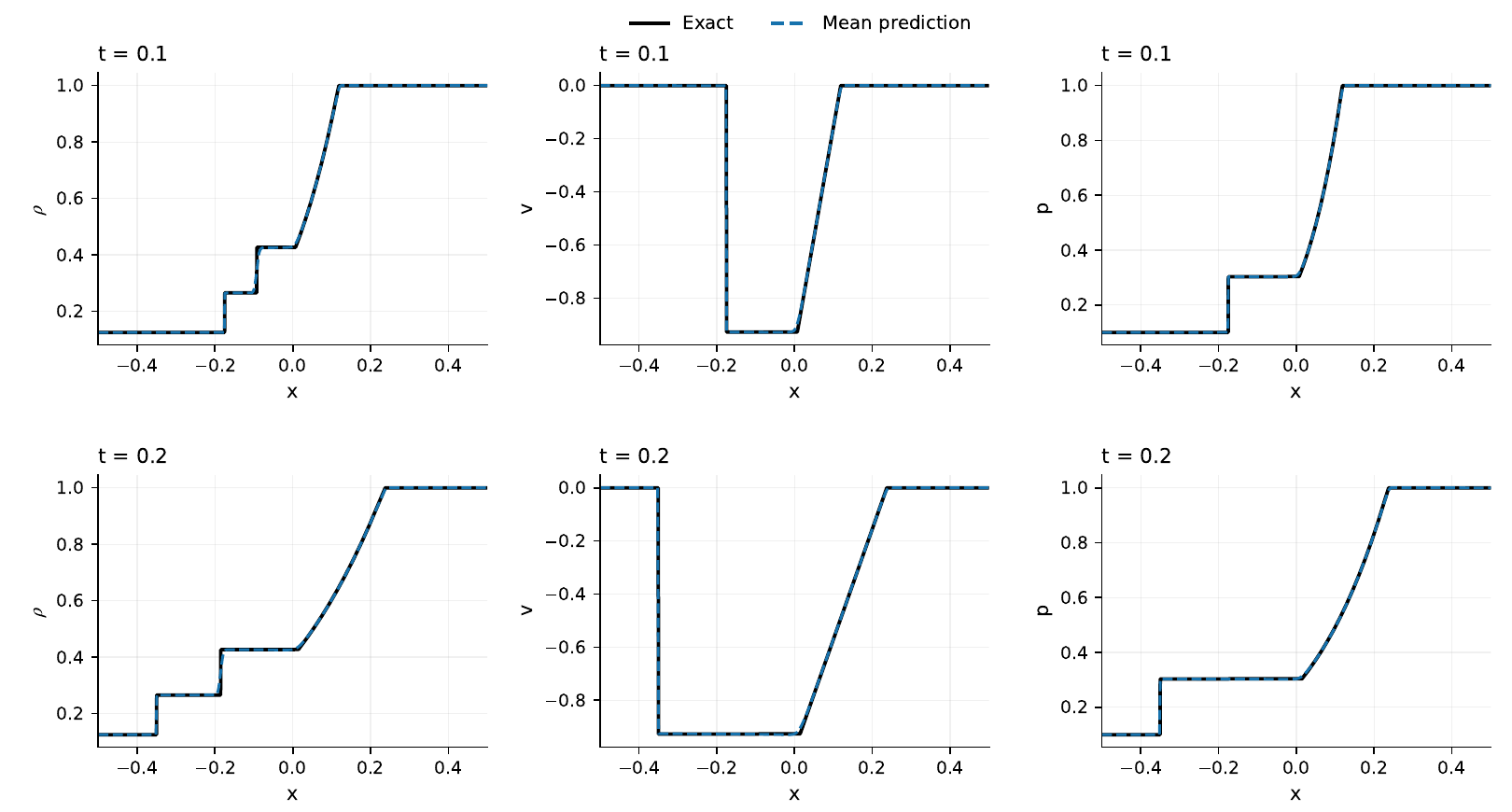}
\caption{Reverse Sod shock tube, obtained by exchanging the left and right states of Sod. The top and bottom rows correspond to $T/2$ and $T=0.2$, and the columns show density, velocity, and pressure. Black solid lines denote the exact solution, and blue dashed lines denote the mean predictions over all ten seeds.}\label{fig:fig06}
\end{figure}

\begin{figure}[!htbp]\centering
\includegraphics[width=\linewidth]{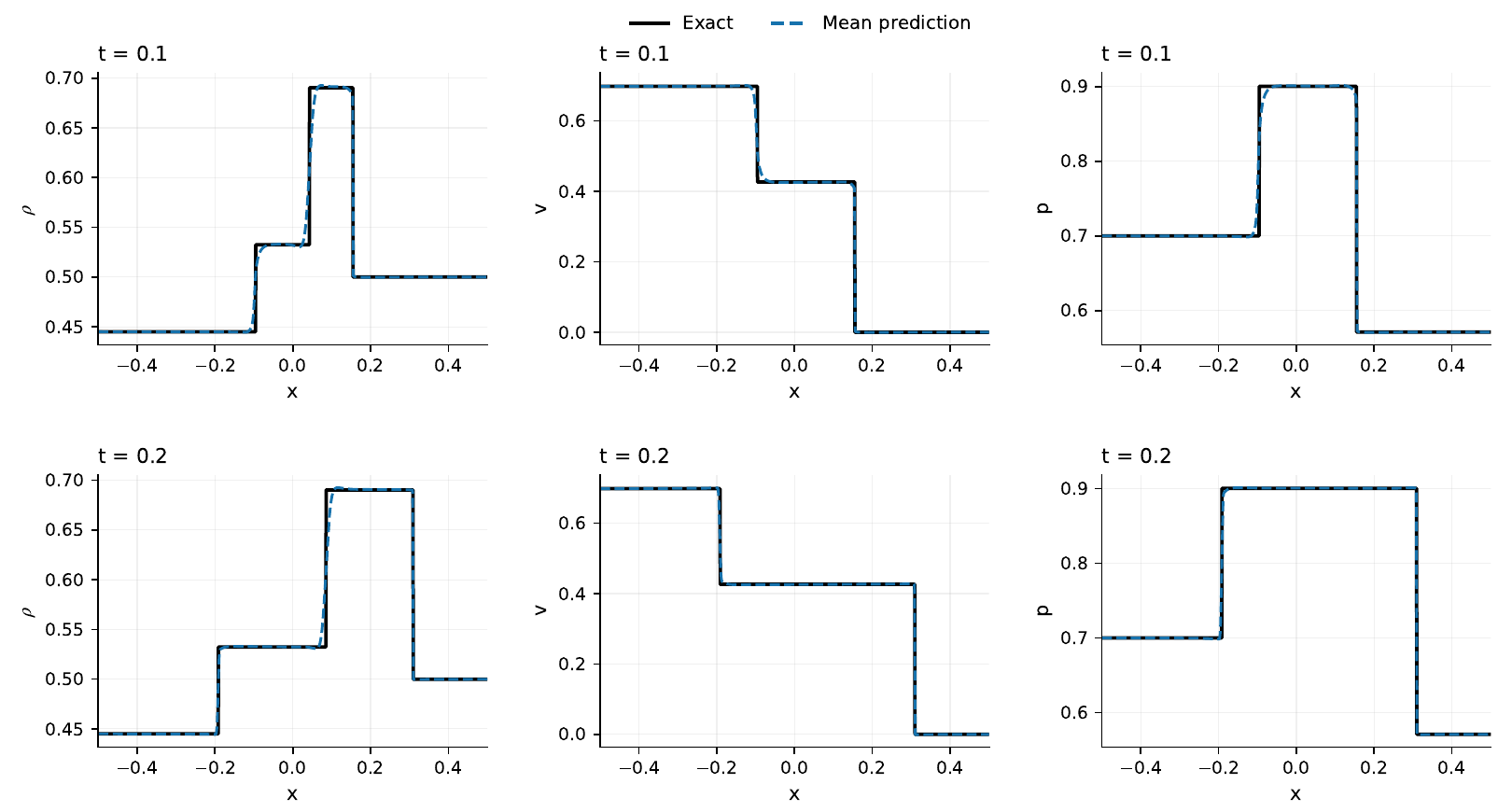}
\caption{Moving-gas Riemann problem. The top and bottom rows correspond to $T/2$ and $T=0.2$, and the columns show density, velocity, and pressure. Black solid lines denote the exact solution, and blue dashed lines denote the mean predictions over all ten seeds.}\label{fig:fig07}
\end{figure}

\begin{figure}[!htbp]\centering
\includegraphics[width=\linewidth]{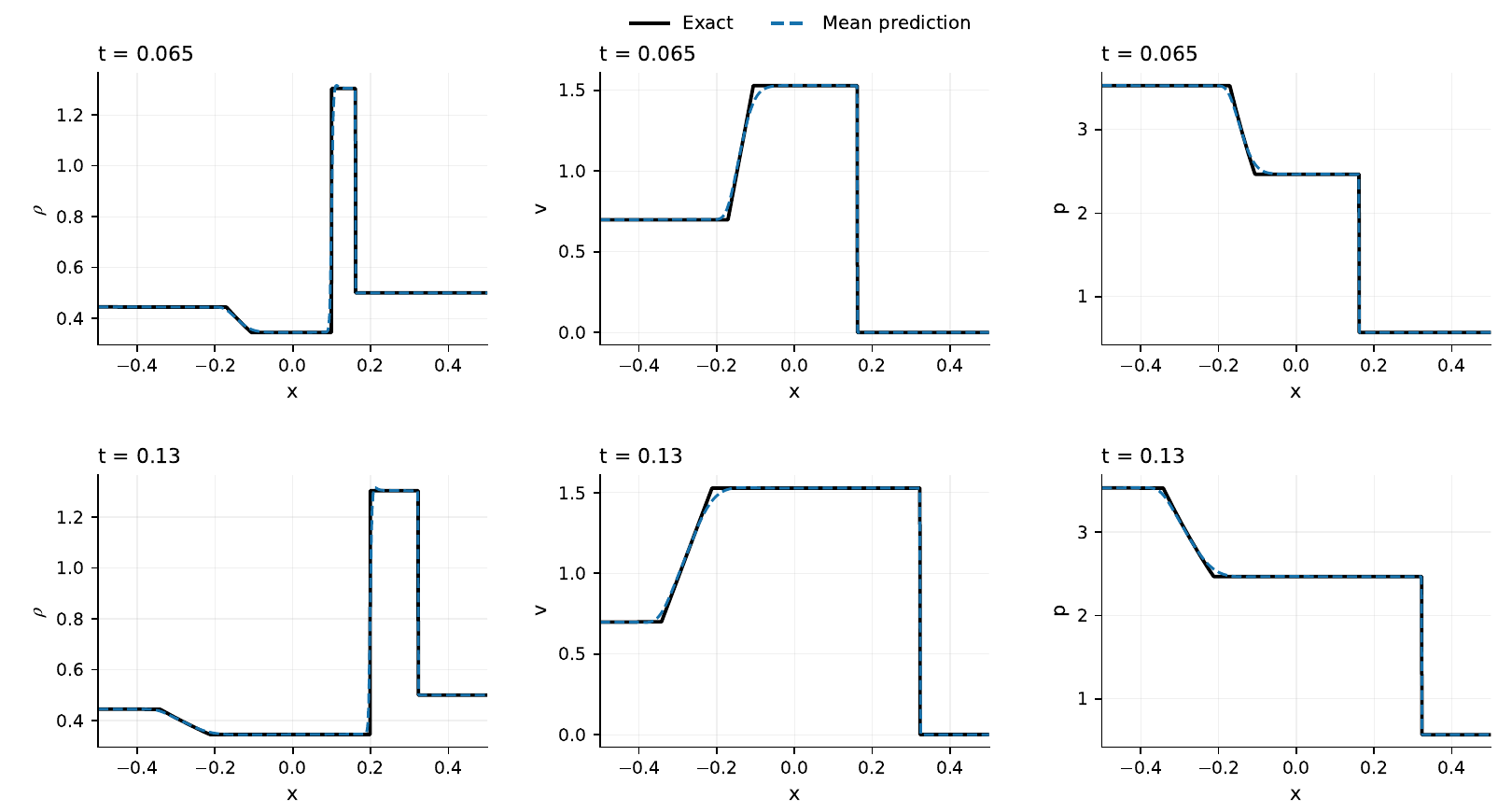}
\caption{Lax Riemann problem. The top and bottom rows correspond to $T/2$ and $T=0.13$, and the columns show density, velocity, and pressure. Black solid lines denote the exact solution, and blue dashed lines denote the mean predictions over all ten seeds.}\label{fig:fig08}
\end{figure}

To examine contact transport separately, we prescribe $v_L=v_R=1$ and $p_L=p_R=0.5$, with left and right densities of either $(1,0.25)$ or $(0.25,1)$. The exact solution is a density step moving at the common velocity; pressure and velocity remain constant. Figure~9 shows the results for both density arrangements.

In both cases, the exact contact position is $x=0.2$ at $T=0.2$. The maximum position error in each of the ten runs is below $8.0\times10^{-5}$. The largest terminal deviations from constant velocity and pressure are $1.33\times10^{-4}$ and $6.4\times10^{-5}$, respectively. For seed~2, the maximum density overshoots are approximately 0.0533 and 0.0686 in the two groups. As noted above, Fig.~9 averages the other four runs, but Table~3 includes all five in each group.

\begin{figure}[!htbp]\centering
\includegraphics[width=\linewidth]{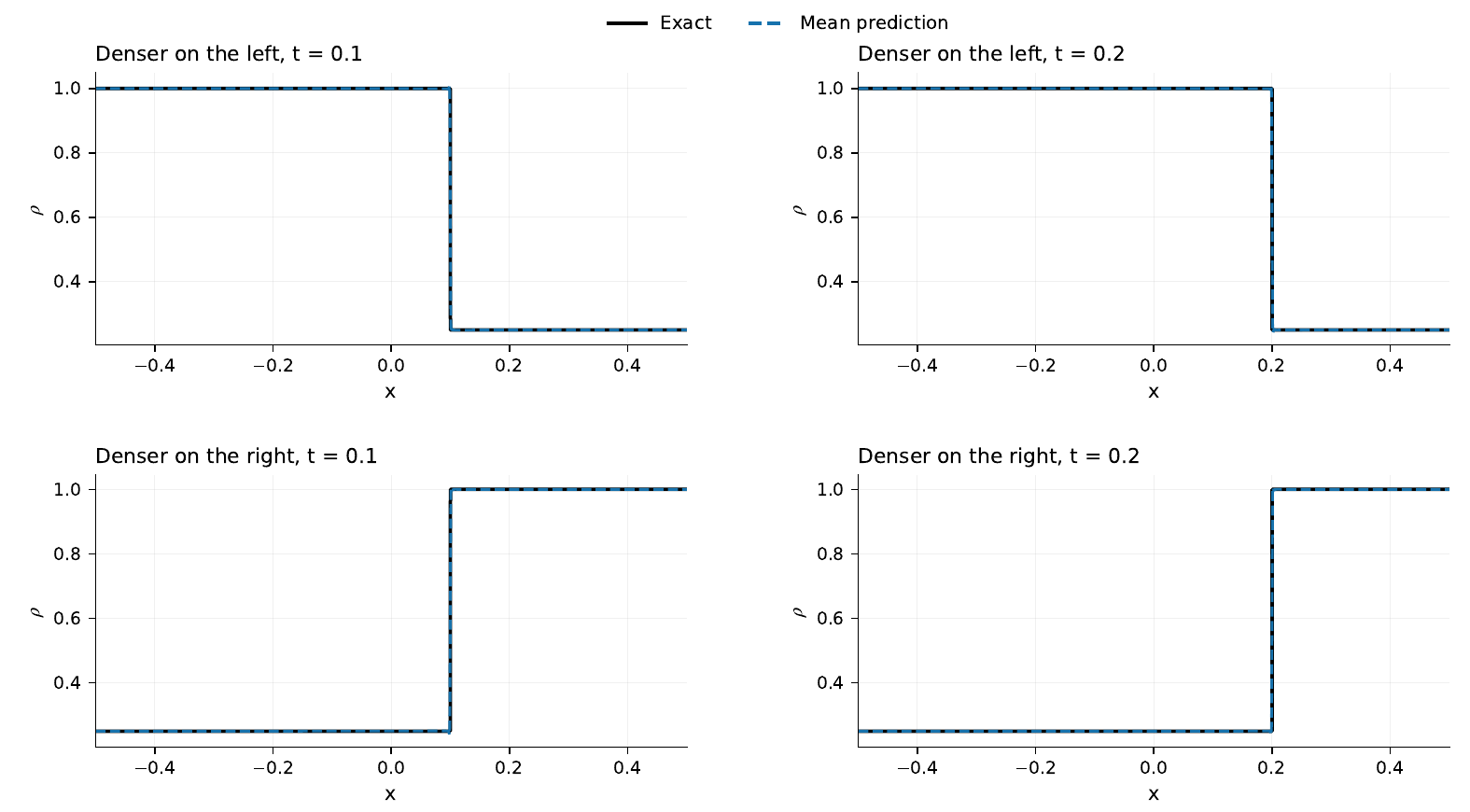}
\caption{Translation of two isolated contact discontinuities. The top row has higher density on the left, and the bottom row has higher density on the right; the columns correspond to $T/2$ and $T=0.2$. Black solid lines denote the exact solution, and blue dashed lines denote the mean predictions of seeds 1, 3, 4, and 5 in each group. Table~3 includes all five seeds.}\label{fig:fig09}
\end{figure}

\subsection{Two-dimensional Burgers equation}\label{sec:burgers2d}

On the unit square, consider $F_1(u)=F_2(u)=u^2/2$, with $G=(-u_x-u_y)_+$ and $T=0.5$. Setting $s=(x+y)/2$ and $r=x-y$ reduces the equation at fixed $r$ to $u_t+(u^2/2)_s=0$, which is used to compute the reference solution. The planar shock initially lies at $x=0.5$, with left and right states $-0.2$ and $-1$. The four-shock problem has its quadrant junction at $(0.5,0.5)$ and initial states $(1,0.5,0.3,-0.5)$ in the lower-left, lower-right, upper-left, and upper-right quadrants, respectively.

\begin{table}[!htbp]\centering
\caption{Individual results for the two-dimensional Burgers problems at $T=0.5$, all using seed~1. Error columns are percentages, obtained by multiplying relative errors by 100. The two planar-shock rows evaluate the same model with different quadratures; the coarse uniform grid does not adequately resolve the transition layer.}\label{tab:4}
\small\setlength{\tabcolsep}{3pt}\renewcommand{\arraystretch}{1.18}
\begin{tabular*}{\linewidth}{@{\extracolsep{\fill}}llcc@{}}\toprule
Problem & Evaluation & $E_1$ (\%) & $E_2$ (\%)\\\midrule
Planar shock & \shortstack[l]{$400^2$ cell centers\\Unresolved transition layer} & 0.00382 & 0.00439\\
Planar shock & \shortstack[l]{$3248\times64$\\Locally refined quadrature} & 0.03140 & 0.82068\\
Four shocks & $400^2$ cell centers & 0.672 & 5.231\\
\bottomrule\end{tabular*}
\end{table}

The two-dimensional network has six hidden layers of width 30 and two potential outputs, and is trained with 3000 Adam steps and 2500 self-scaled BFGS steps. The planar- and four-shock problems use 6000 and 30000 interior collocation points, respectively. Boundary potentials are determined by integrating flux traces from the one-dimensional boundary problems.

The exact planar shock follows $x_s=0.5-0.6t$, reaching $x=0.2$ at the terminal time (Fig.~S5). The median diagnostic layer width for seed~1 is 0.000625, well below the spacing 0.0025 of the $400^2$ cell-center grid. At the terminal time, the shock lies between adjacent sampling points. The resulting coarse-grid error $E_2=0.00439\%$ substantially underestimates the continuous-field error. Local quadrature grids of $812\times16$, $1624\times32$, and $3248\times64$, with integration split at the exact jump, give $E_2=0.82005\%$, $0.82055\%$, and $0.82068\%$, respectively. Table~4 reports both the finest-quadrature result and the coarse-grid value to illustrate this resolution effect.

By $T=0.5$, the four-shock solution contains several front segments and junctions (Fig.~10). For seed~1, $E_1=0.672\%$ and $E_2=5.23\%$. The horizontal and vertical sections are supplemented by oblique sections along $x+y=1.0,1.2,1.3,1.4$, which cross two or three discontinuities. The predicted constant-state regions and the sequence of jumps agree with the reference. Near the junction, however, the section $x+y=1.3$ reveals appreciable broadening and displacement. The median diagnostic layer width is 0.00560, and the maximum position error is 0.00518.

Position and width diagnostics are evaluated at $t/T=0.1,0.25,0.5,0.75,1$, using 4000 cell centers along each of 41 constant-$r$ lines. When adjacent reference samples decrease by more than 5\% of the initial range, their midpoint defines a front position. The nearest descending half-amplitude crossing is then matched within an $x$-distance of 0.05. Two-dimensional error maps use a logarithmic color scale from $10^{-4}$ to 1; error norms are computed from unclipped values.

\begin{figure}[!htbp]\centering
\includegraphics[width=\linewidth]{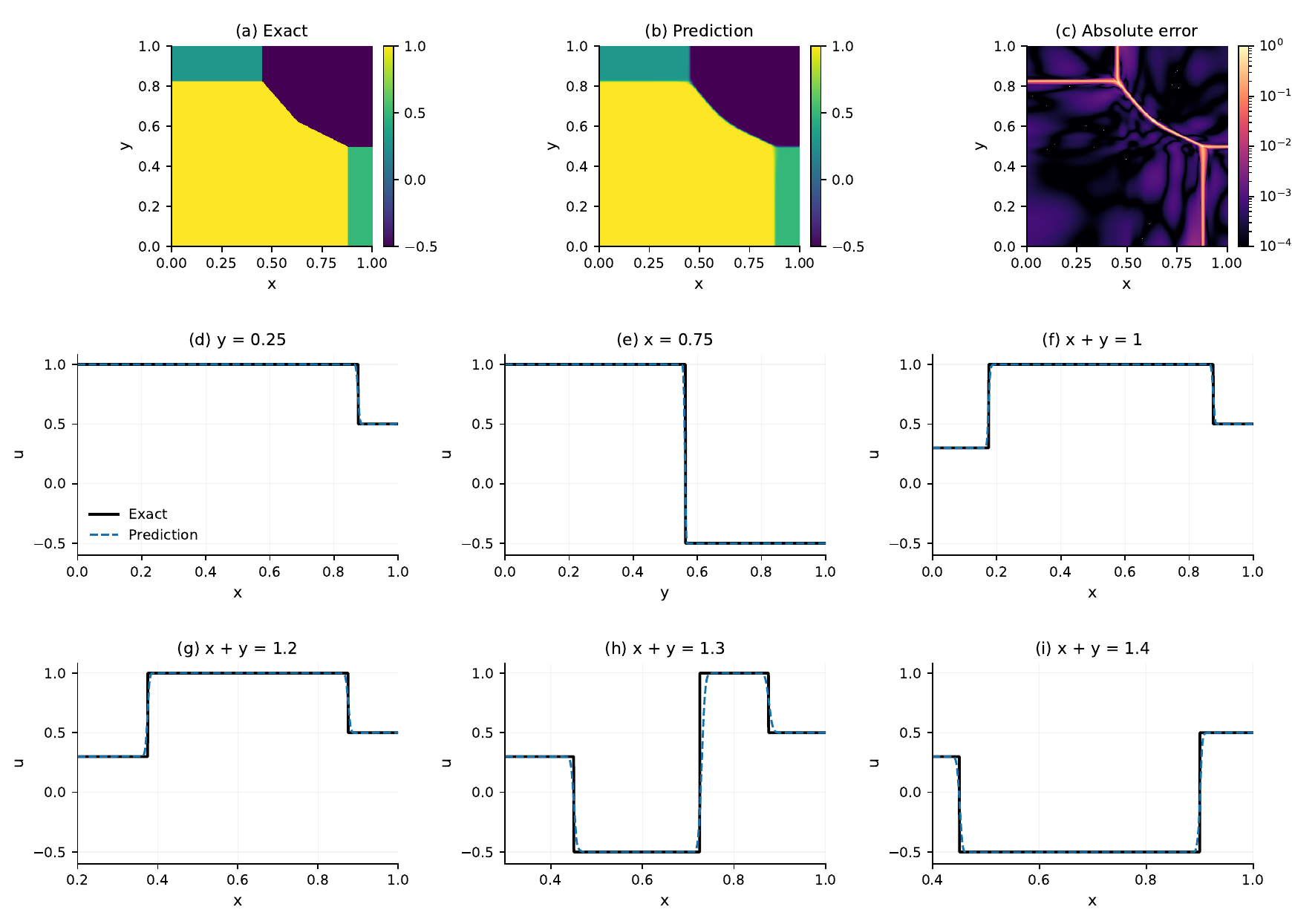}
\caption{Seed~1 results for the four-shock problem at $T=0.5$. Panels (a)--(c) show the exact field, prediction, and absolute error; (d) and (e) show horizontal and vertical sections; and (f)--(i) show oblique sections along $x+y=1.0,1.2,1.3,1.4$. Black solid lines denote the exact solution, and blue dashed lines denote predictions from the same model. Each section is evaluated directly at 3001 points along the line.}\label{fig:fig10}
\end{figure}

%% file: operator_results.tex
\clearpage

\section{Numerical experiments for conditional solution operators}\label{sec:opresults}

\subsection{Problem settings and evaluation}\label{sec:opsettings}

The operator experiments consider three initial-data families for the ideal-gas Euler equations. In all cases, the initial interface lies at $x=0.5$, the right state is $(0.125,0,0.1)$, and the terminal time is $T=0.1$. The low-pressure-ratio family uses the left state $(1,0,p_L)$ with $p_L\in[1,5]$ on $[0,1]$. The higher-pressure-ratio family uses $(2,0,p_L)$ with $p_L\in[50,100]$ on $[-1,2]$. The left-to-right pressure ratios are 10--50 and 500--1000, respectively. Both scalar-input families are evaluated against exact Riemann profiles at 4097 points. The pressure-function family and its reference solutions are defined in Section~5.3.

Within each family, all methods use the same training and test inputs. For each network, we average the relative $L^2$ errors of the three primitive variables, then average over the 100 test inputs. The resulting run-level errors are summarized by their mean and sample standard deviation over three seeds. Table~6 reports a separate ablation with three label counts and three seeds per count; its standard deviation therefore includes variation across both label counts and seeds.

The operator profiles show seed~1 for the low-pressure-ratio family, seed~2 for function inputs, and seed~3 for the higher-pressure-ratio family. The seed in each family is the lowest-numbered seed common to the two better-performing seeds of each method, ranked by terminal refined-grid error. Each figure uses the same seed and input across methods. This selection applies only to profile plots; all three seeds enter the statistical tables and Fig.~14.

\begin{table}[!htbp]\centering
\caption{Terminal mean relative $L^2$ errors (\%) over 100 test initial conditions, reported as mean $\pm$ standard deviation over three seeds. Scalar inputs are evaluated at 4097 points and function inputs at 6145 points. The no-residual model retains the same initial-data extension, warm-up, boundary potentials, and data loss as the dual-residual model.}\label{tab:5}
\small\setlength{\tabcolsep}{3pt}\renewcommand{\arraystretch}{1.18}
\begin{tabular*}{\linewidth}{@{\extracolsep{\fill}}lcccc@{}}\toprule
Input family & \shortstack[l]{Labeled initial\\conditions} & Dual residual & No residual & \shortstack[l]{Local\\RiemannONet}\\\midrule
\shortstack[l]{Low pressure\\ratio} & 25 & $1.21 \pm 0.108$ & $1.38 \pm 0.0513$ & $6.88 \pm 0.57$\\
 & 100 & $0.983 \pm 0.193$ & $2.16 \pm 2.37$ & $5.37 \pm 1.01$\\
 & 400 & $0.94 \pm 0.054$ & $0.923 \pm 0.791$ & $4.6 \pm 0.766$\\
\addlinespace[3pt]
Function input & 25 & $2.76 \pm 0.935$ & $2.06 \pm 0.384$ & $16.6 \pm 0.987$\\
 & 100 & $1.4 \pm 0.211$ & $1.33 \pm 0.342$ & $4.93 \pm 0.584$\\
 & 400 & $1.12 \pm 0.204$ & $0.722 \pm 0.252$ & $3.35 \pm 0.399$\\
\addlinespace[3pt]
\shortstack[l]{Higher pressure\\ratio} & 25 & $1.31 \pm 0.508$ & $0.941 \pm 0.278$ & $6.3 \pm 0.761$\\
 & 100 & $2.22 \pm 2.31$ & $0.664 \pm 0.0831$ & $5.02 \pm 1.06$\\
 & 400 & $1.06 \pm 0.234$ & $0.279 \pm 0.0689$ & $4.79 \pm 0.791$\\
\bottomrule\end{tabular*}
\end{table}

\subsection{Scalar-pressure inputs}\label{sec:scalar}

Figure~11 shows predictions for the low-pressure-ratio family. Both the right-moving shock and the contact travel farther as $p_L$ increases. With 25 labeled initial conditions, the mean shock-position errors of the dual-residual models lie between 0.0003 and 0.0005 across seeds, and the mean 10\%--90\% layer widths between 0.0007 and 0.0009. The corresponding ranges for the RiemannONet baseline are 0.0017--0.0018 and 0.0040--0.0048. At this label count, each of the three dual-residual models gives a lower terminal error on every one of the 100 test inputs.

The evaluation grid matters in this comparison. At the 200 label locations, the terminal errors of the 25-label dual-residual models (0.52\%--0.74\%) are close to those of the 400-label RiemannONet (0.56\%--0.83\%). The latter errors increase on the 4097-point grid because the finer grid resolves oscillations between label locations. All entries in Table~5 are based on the refined grid.

The enlarged profiles in Fig.~11 show that both potential-based models follow the exact terminal solution, with the main differences occurring near the contact and the shock. The RiemannONet results are for the local sine-Rowdy implementation described in Section~3.2, rather than the recommended public configuration for these benchmarks. This distinction is important when interpreting the comparison.

\begin{figure}[!htbp]\centering
\includegraphics[width=\linewidth]{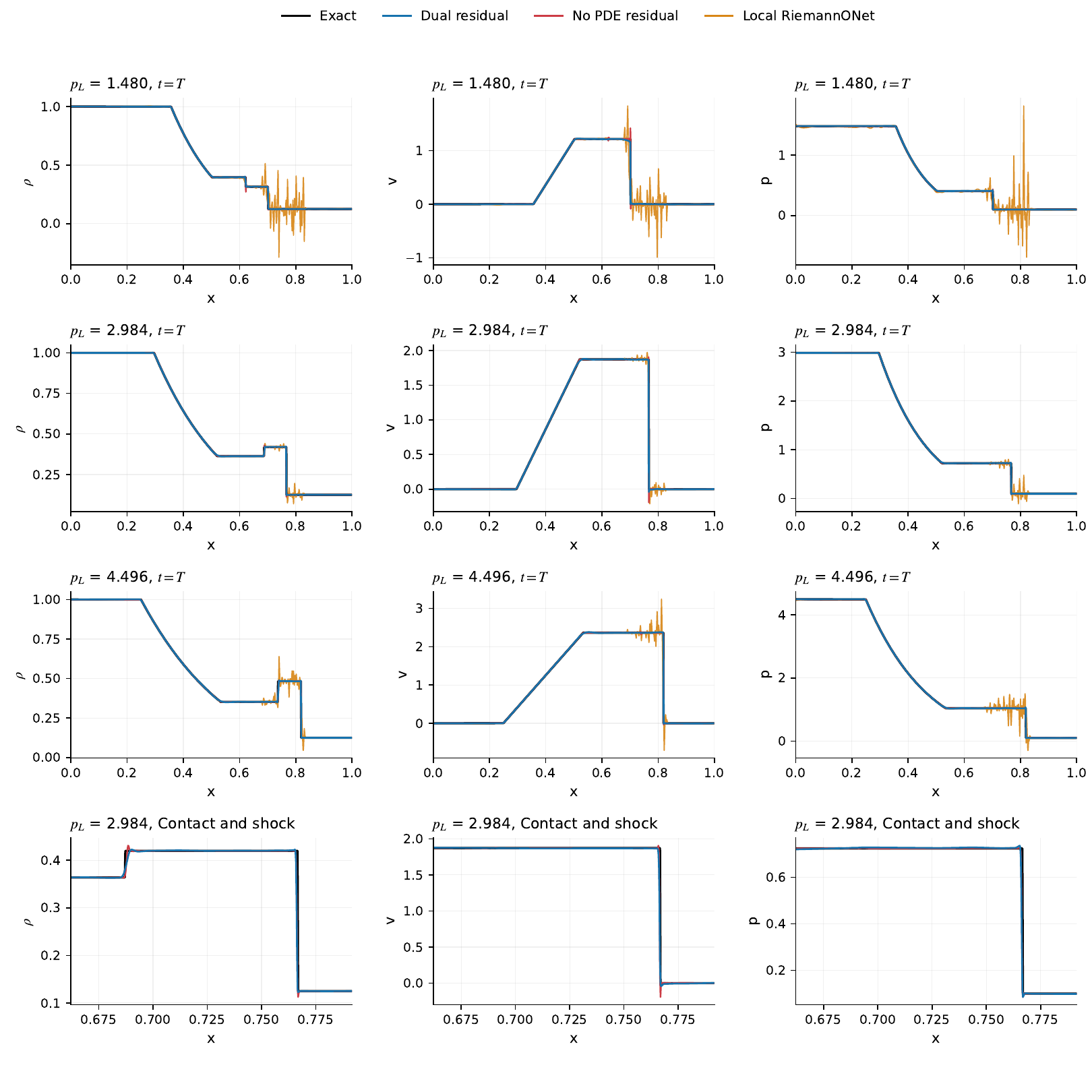}
\caption{Terminal predictions for the low-pressure-ratio family, using 25 labeled initial conditions and seed~1. The first three rows correspond to $p_L\approx1.48,2.98,4.50$, and the columns show density, velocity, and pressure. The fourth row gives local sections for the middle-pressure case, comparing only the dual-residual and no-residual models. Black solid lines denote the exact solution.}\label{fig:fig11}
\end{figure}

Figure~12 compares the two potential-based models in more detail at $p_L\approx2.984$. Their relative errors, averaged over the three variables, are 1.39\% (dual residual) and 1.40\% (no PDE residuals). Both capture the rarefaction and the principal discontinuities. The enlarged plots show greater overshoot near the shock in velocity for the no-residual model; neither model eliminates the finite width of the contact layer.

\begin{figure}[!htbp]\centering
\includegraphics[width=\linewidth]{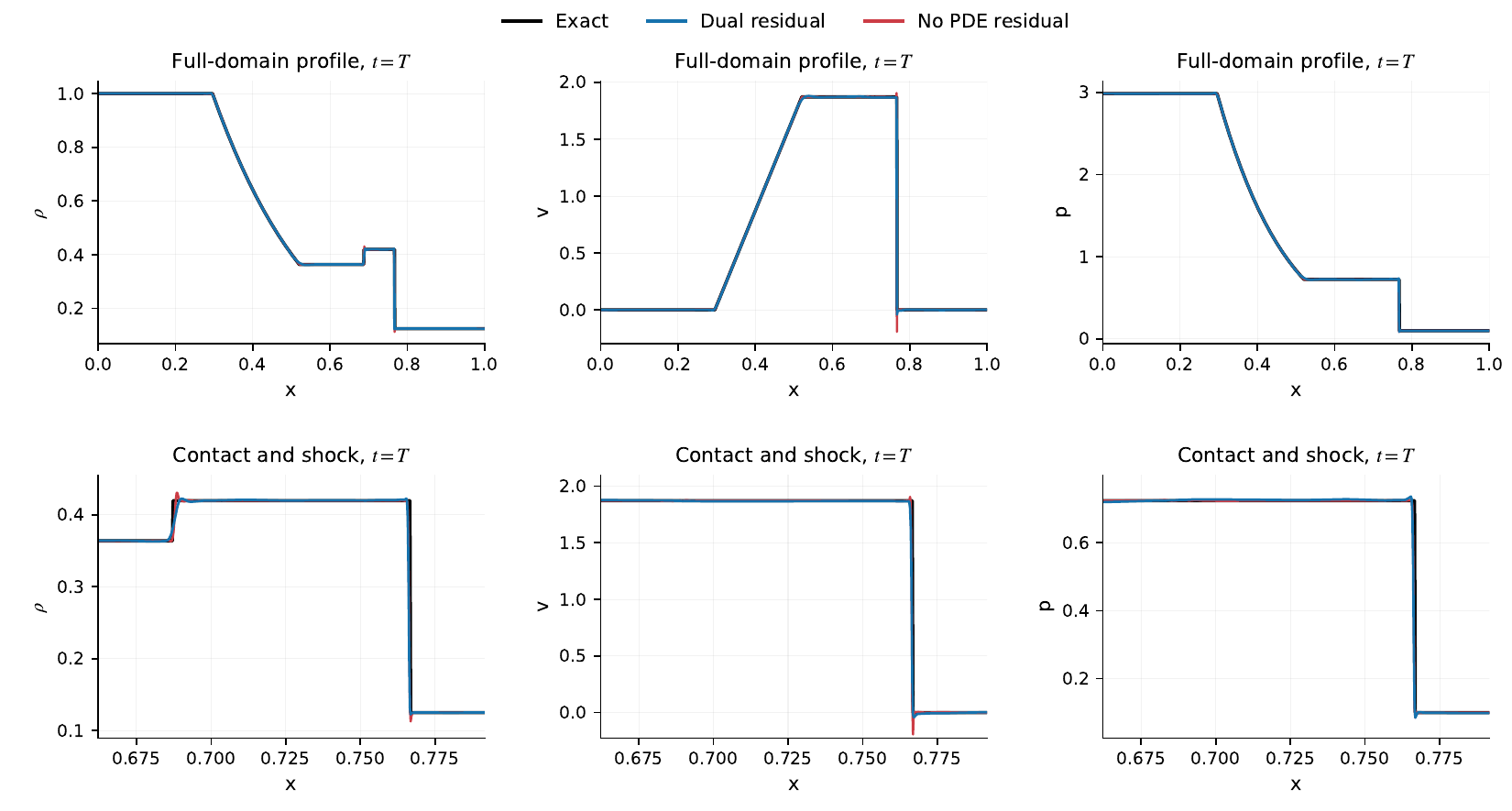}
\caption{Terminal profile comparison at $p_L\approx2.984$, using 25 labeled initial conditions and seed~1. The top row gives full-domain profiles at $T=0.1$, and the bottom row gives local sections near the contact and shock; the columns show density, velocity, and pressure. Local vertical limits extend the exact-solution range by 12\%. Black, blue, and red denote the exact solution, dual-residual model, and no-PDE-residual model, respectively. The evaluation grid contains 4097 points.}\label{fig:fig12}
\end{figure}

\begin{table}[!htbp]\centering
\caption{Loss ablations for the low-pressure-ratio family. Each configuration includes budgets of 25, 100, and 400 labels and three seeds, giving nine runs. Severe overshoot means that the 95th percentile of normalized pressure overshoot over test profiles exceeds 0.5. The final column reports the mean $\pm$ standard deviation of the error at the unsupervised time $T/2$.}\label{tab:6}
\small\setlength{\tabcolsep}{3pt}\renewcommand{\arraystretch}{1.18}
\begin{tabular*}{\linewidth}{@{\extracolsep{\fill}}llccc@{}}\toprule
Retained residuals & \shortstack[l]{Gradient\\downweighting} & \shortstack[l]{Non-positive\\states} & \shortstack[l]{Severe\\overshoot} & $E_2(T/2)$ (\%)\\\midrule
Potential and strong & Both & 1/9 & 0/9 & $4.09 \pm 1.18$\\
Strong only & Yes & 3/9 & 0/9 & $5.21 \pm 1.39$\\
Potential and strong & Neither & 3/9 & 0/9 & $8.98 \pm 0.464$\\
Strong only & No & 0/9 & 0/9 & $8.72 \pm 0.591$\\
Potential only & Yes & 7/9 & 6/9 & $6.88 \pm 4.31$\\
Potential only & No & 5/9 & 4/9 & $5.23 \pm 3.02$\\
\bottomrule\end{tabular*}
\end{table}

Table~6 separates the effects of residual choice and gradient downweighting. The complete model gives a mean half-time error of 4.09\%, with no severe overshoots in the nine runs. Removing both downweighting factors increases the error to 8.98\%. Six of the nine runs with only the downweighted potential residual exhibit severe overshoot. The downweighted strong-form residual alone avoids severe overshoot but gives a larger mean error than the combined loss. These results suggest complementary roles for the two residuals in constraining local variation and time evolution. One complete-model run in this ablation reaches a minimum pressure of approximately $-0.009$; by contrast, all nine complete models in the main low-pressure-ratio experiments pass the refined-grid terminal positivity checks. The two experiments use independently trained models. Example profiles are given in Fig.~S6.

\subsection{Pressure-function inputs}\label{sec:function}

On $[-0.5,1]$, the left density is 1 and the left velocity is zero. The initial pressure is $a$ for $x<0$, and, between 0 and 0.5, it is

\[p_{L}(x) = a + (c - a)\xi + 4d\,\xi(1 - \xi),\quad\quad\xi = 2x,\quad a,c \in \lbrack 1,5\rbrack,\quad d \in \lbrack - 0.5,0.5\rbrack.\]

We draw the coefficients independently from uniform distributions to obtain 400 training, 80 validation, and 100 test functions. The network receives pressure values at 16 equally spaced sensors on $[0,0.5]$. The initial potential uses the primitive of a cubic spline through these readings; for the quadratic family considered here, this interpolation recovers the input pressure exactly. Test coefficients remain within the training parameter ranges. For a label count $n$, all methods receive the first $n$ functions from the common training pool.

Reference solutions are computed by a second-order MUSCL--Hancock finite-volume scheme in Lagrangian mass coordinates on 3000 cells. At output, the main shock is reconstructed as a step using an equal-area condition. For $(a,c,d)=(1,5,0.5),(5,1,-0.5),(3,3,0.5)$, refining to 6000 cells gives mean relative differences over the three variables of 0.0676\%--0.0851\% in the terminal 6145-point profiles. Training labels use 301 terminal points. Testing uses a 6145-point grid, with the two space--time potential models also evaluated at $T/4,T/2,3T/4$. The RiemannONet baseline provides only terminal predictions.

Figure~13 shows predictions for an increasing pressure function; decreasing and peaked inputs are shown in Figs.~S7 and S8. The three input types share a trained network, with only the sensor readings changed. Terminal errors of the dual-residual model are 3.25\%, 3.00\%, and 1.81\%, respectively, compared with 5.66\%, 11.12\%, and 16.85\% for RiemannONet. The examples are selected by the proximity of their coefficients to prescribed function shapes and all use seed~2. Another increasing input is shown in Fig.~S9.

In the 25-label experiments, terminal errors are 1.63\%--2.38\% for the no-residual model and 2.12\%--3.83\% for the dual-residual model. At the unsupervised time $T/2$, however, their errors are 33\%--56\% and 5.4\%--7.4\%, respectively. The principal effect of the physical residuals is therefore to improve predictions within the time interval, rather than to reduce terminal fitting errors in every setting. At the 301 label locations, the 25-label complete model has errors of 1.22\%--1.95\%, below the 2.42\%--2.54\% of the 400-label RiemannONet. Their refined-grid error ranges overlap. Table~7 gives paired comparisons for each test input with 25 labels. One complete model produces negative pressure on one test function, with a minimum of $-4.35$.

\begin{figure}[!htbp]\centering
\includegraphics[width=\linewidth]{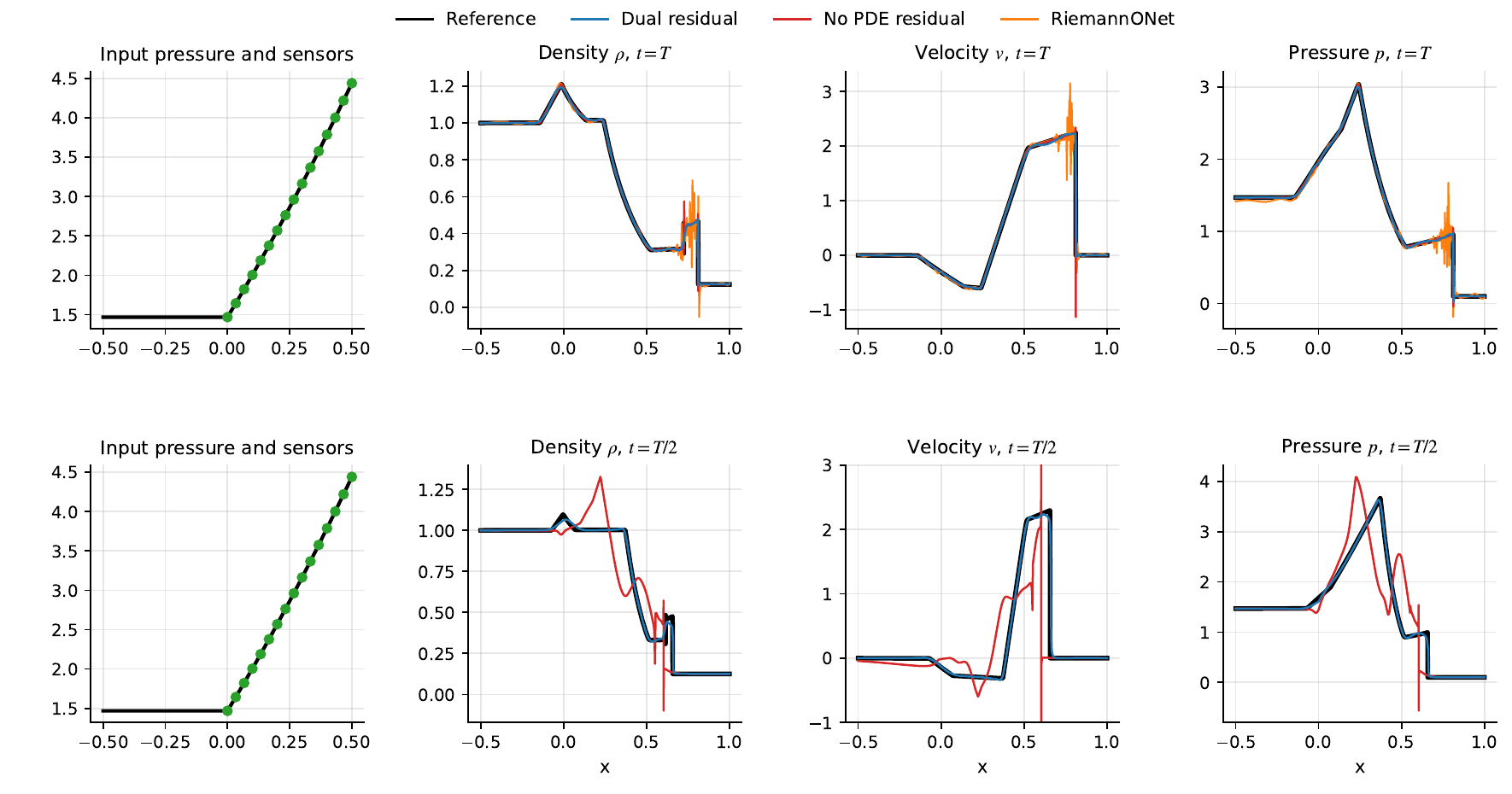}
\caption{Conditional predictions for an increasing pressure input, using test function~50, 25 training functions, and seed~2. The first column shows the input pressure and 16 sensors; the remaining columns show density, velocity, and pressure. The top and bottom rows correspond to $T$ and $T/2$. Black solid lines denote the reference solution. The bottom-row velocity axis is restricted to $[-1,3]$; error metrics use unclipped predictions. RiemannONet provides only terminal profiles.}\label{fig:fig13}
\end{figure}

\begin{table}[!htbp]\centering
\caption{Paired comparisons by test input with 25 labels. Entries correspond to the three seeds in order and give terminal mean relative $L^2$ errors (\%), together with the number of test inputs on which the dual-residual error is lower than that of the RiemannONet baseline used here. Each seed uses the same 100 test inputs.}\label{tab:7}
\small\setlength{\tabcolsep}{3pt}\renewcommand{\arraystretch}{1.18}
\begin{tabular*}{\linewidth}{@{\extracolsep{\fill}}lccc@{}}\toprule
Input family & Dual residual (\%) & RiemannONet (\%) & \shortstack[l]{Lower-error inputs\\(out of 100)}\\\midrule
\shortstack[l]{Low pressure\\ratio} & 1.16, 1.13, 1.33 & 6.32, 7.46, 6.87 & 100, 100, 100\\
Function input & 3.83, 2.33, 2.12 & 17.53, 15.57, 16.72 & 99, 100, 100\\
\bottomrule\end{tabular*}
\end{table}

\subsection{Physical admissibility at higher pressure ratios}\label{sec:admissibility}

Results for the higher-pressure-ratio family are given in Table~5 and Fig.~S10. With 25 labeled initial conditions, terminal errors are 0.89\%--1.87\% for the complete model and 5.73\%--7.16\% for the RiemannONet baseline. Nevertheless, 93\%--100\% of the complete model's test profiles contain non-positive density or pressure; minimum pressures across the three seeds range from $-4.43$ to $-0.88$. The baseline has non-positive states in 99\%--100\% of its profiles. Increasing the label count to 400 gives complete-model terminal errors of 0.84\%--1.31\%, but non-positive states remain detectable in every test profile. The integral-potential dual-residual construction therefore does not yet guarantee physical admissibility in high-pressure-ratio flows.

Figure~14 summarizes 81 combinations of three input families, three methods, three label budgets, and three seeds. Differences between label-grid and refined-grid errors reflect resolution effects near discontinuities, while the fraction of non-positive profiles identifies physical-constraint deficiencies not captured by the mean error.

\begin{figure}[!htbp]\centering
\includegraphics[width=\linewidth]{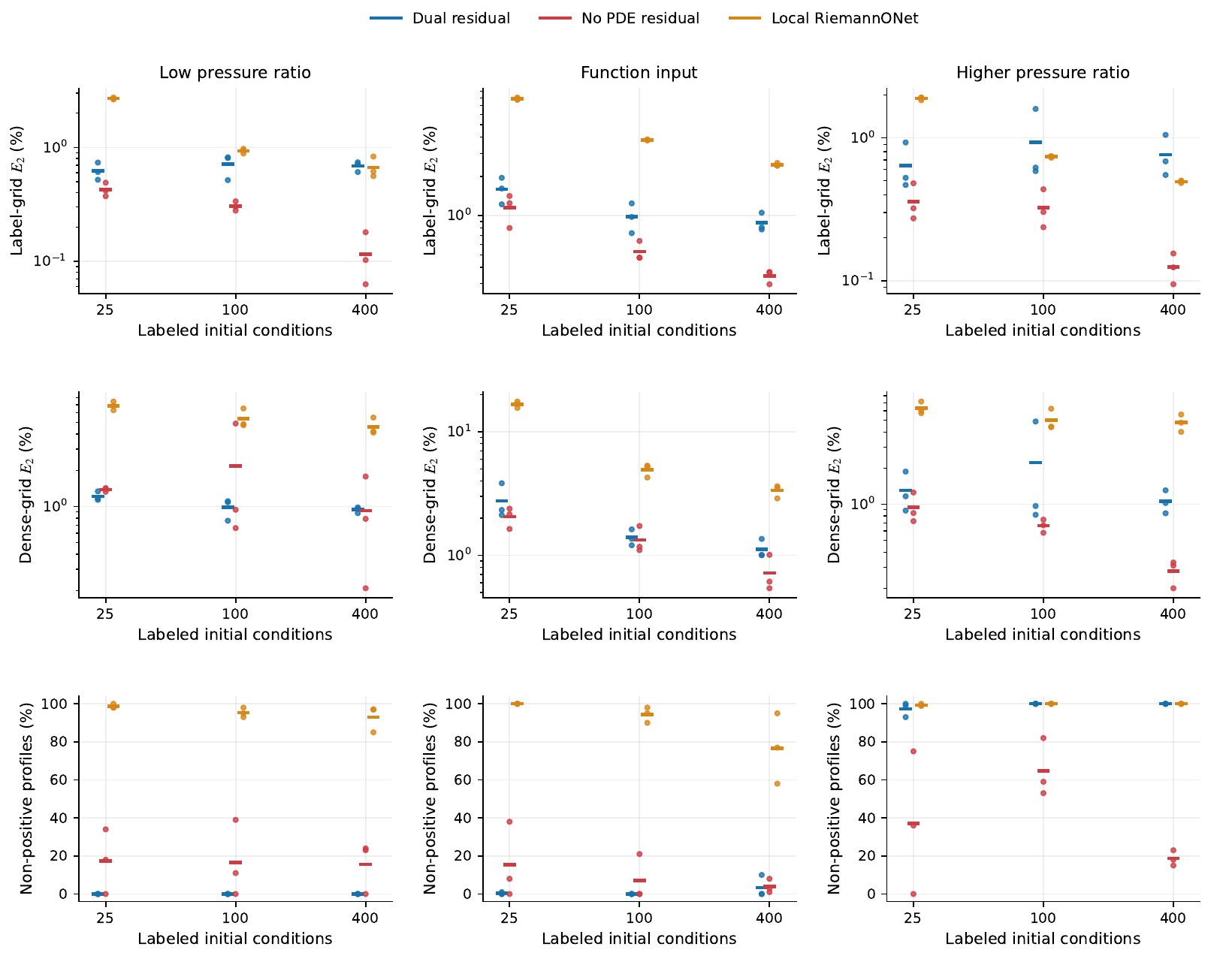}
\caption{Effect of the number of labeled initial conditions. Columns correspond to the low-pressure-ratio, function-input, and higher-pressure-ratio families. Rows show label-grid errors, dense-grid errors, and the percentage of test profiles with non-positive states. Each group includes dots for all three seeds, with short horizontal lines indicating arithmetic means; coincident points may overlap.}\label{fig:fig14}
\end{figure}

%% file: discussion.tex
\section{Discussion and conclusions}\label{sec:conclusions}

We have incorporated an integral-potential representation into gwPINN and extended the resulting method to conditional solution-operator learning. The potential residual provides a constraint on interval conservation, while the strong-form residual penalizes its spatial variation. Each residual is downweighted in compression regions. In two paired moving-shock tests for the Burgers equation, adding the potential residual substantially reduces the position error, including a case in which the layer widths are nearly identical. Euler Riemann problems and two-dimensional Burgers tests illustrate the performance for other wave types and interactions.

The potential equation alone does not provide sufficient control of local oscillations in the experiments. In the independent nine-run low-pressure-ratio operator ablation, severe overshoot occurs in six runs with only the downweighted potential residual, but in none of the runs using both downweighted residuals. For the finite networks tested here, it is useful to impose both the cumulative transport constraint and its local differential counterpart.

For operator learning, the shared network, initial-data encoding, and joint collocation sampling allow the same physical losses to be applied over a family of initial conditions. With 25 terminal labels, the method gives smaller terminal errors than our RiemannONet baseline for every low-pressure-ratio test input and nearly every pressure-function input. The comparison with the same architecture trained without PDE residuals is especially relevant: for pressure-function inputs, the control can fit the terminal profiles as well as or better than the complete model, yet its errors at unsupervised intermediate times are much larger. In this setting, the principal benefit of the PDE residuals is to constrain the evolution between the initial and terminal times.

There are several limitations. The occurrence of negative pressure at higher pressure ratios points to the need for explicit positivity constraints or other treatments of the admissible set \citep{r07}. The function inputs explored here have only three degrees of freedom, and the two-dimensional Burgers tests use identical directional fluxes. More general function families and multidimensional systems still need to be investigated. The planar-shock experiment also shows that a coarse evaluation grid can substantially underestimate the error in a thin transition layer. Differences in baseline implementation, the small number of training replicates, and the use of some test metrics during development limit the conclusions that can be drawn from the comparisons \citep{r51}. Further work will consider positivity, entropy constraints, and more general boundary conditions for the integral-potential formulation.

\paragraph{Data availability.} The two scalar-pressure benchmarks use the public RiemannONets data \citep{r46}. Pressure-function data are generated by the finite-volume method described here. Numerical records include training configurations, run-level errors, per-test-input results, and identifiers of the models shown in the figures.

\paragraph{Declaration of generative AI use.} The authors used OpenAI Codex to assist in modifying their existing code and implementing the solver programs and experimental scripts, and to polish the wording.

%% file: references.tex
\clearpage\input{main.bbl}

%% file: appendix_a.tex
\clearpage\appendix
\section{Comparison of sampling and temporal weighting}\label{app:sampling}

Table~8 compares the original configuration with a control using uniform collocation and no time window. Since the control removes adaptive sampling and early-time downweighting together, the comparison measures their combined effect. Initial data, seeds, and residual coefficients are unchanged within each group, as is the restart procedure for time-slab runs. Section~4.3 reports a separate comparison of collocation rules with a common starting point and optimization budget.

\begin{table}[!htbp]\centering
\caption{Terminal relative $L^2$ errors (\%) for the original configuration and the uniformly sampled control without a time window. One-dimensional cases report mean $\pm$ standard deviation over all seeds; the two-dimensional four-shock case reports a single run with seed~1.}\label{tab:8}
\small\setlength{\tabcolsep}{3pt}\renewcommand{\arraystretch}{1.18}
\begin{tabular*}{\linewidth}{@{\extracolsep{\fill}}llccc@{}}\toprule
Equation & Initial data & Seeds & Original (\%) & \shortstack[l]{Uniform sampling\\No time window (\%)}\\\midrule
1D Burgers & Moving shock & 5 & $0.646 \pm 0.0929$ & $1.3 \pm 0.188$\\
 & Stationary shock & 5 & $1.27 \pm 0.128$ & $2.09 \pm 0.088$\\
 & Centered rarefaction & 5 & $0.0739 \pm 0.00649$ & $0.179 \pm 0.084$\\
 & Shock merging & 5 & $1.51 \pm 0.216$ & $1.97 \pm 0.236$\\
 & Square pulse (slabs) & 5 & $2.56 \pm 0.313$ & $2.99 \pm 0.77$\\
 & Sinusoid (slabs) & 5 & $7.03 \pm 0.59$ & $8.37 \pm 0.792$\\
\addlinespace[3pt]
1D Euler & Sod & 10 & $1.27 \pm 0.453$ & $2.83 \pm 0.809$\\
 & Reverse Sod & 10 & $1.71 \pm 0.689$ & $2.48 \pm 0.585$\\
 & Moving gas & 10 & $1.38 \pm 0.166$ & $1.66 \pm 0.175$\\
 & Lax & 10 & $2.59 \pm 0.335$ & $4.04 \pm 1.47$\\
 & \shortstack[l]{Contact\\(denser left)} & 5 & $0.314 \pm 0.0266$ & $0.573 \pm 0.0756$\\
 & \shortstack[l]{Contact\\(denser right)} & 5 & $0.442 \pm 0.0412$ & $0.739 \pm 0.0706$\\
\addlinespace[3pt]
2D Burgers & Four shocks & 1 & 5.231 & 6.458\\
\addlinespace[3pt]
\shortstack[l]{Euler\\Two interfaces} & Colliding shock tubes & 5 & $5.02 \pm 0.695$ & $10.6 \pm 1.61$\\
 & Shock--contact interaction & 5 & $4.08 \pm 0.808$ & $8.41 \pm 3.3$\\
\bottomrule\end{tabular*}
\end{table}

The final two Euler problems each start from two interfaces and are evaluated against fine-grid finite-volume references after interaction. In the colliding-shock-tube problem, a central state $(0.125,0,0.1)$ on $(-0.2,0.2)$ separates outer states $(1,0,1)$, and $T=0.2$. For the shock--contact interaction, the three states are $(1,0,1)$, $(0.125,0,0.1)$, and $(0.5,0,0.1)$, with interfaces at $-0.15$ and $0.15$, and $T=0.25$. In the five uniformly sampled colliding-shock-tube runs, the restart-detection procedure does not trigger an actual time subdivision.

%% file: appendix_b.tex
\clearpage\section{Supplementary full-domain profiles}\label{app:profiles}
\setcounter{figure}{0}\renewcommand{\thefigure}{S\arabic{figure}}

\begin{figure}[!htbp]\centering
\includegraphics[width=\linewidth]{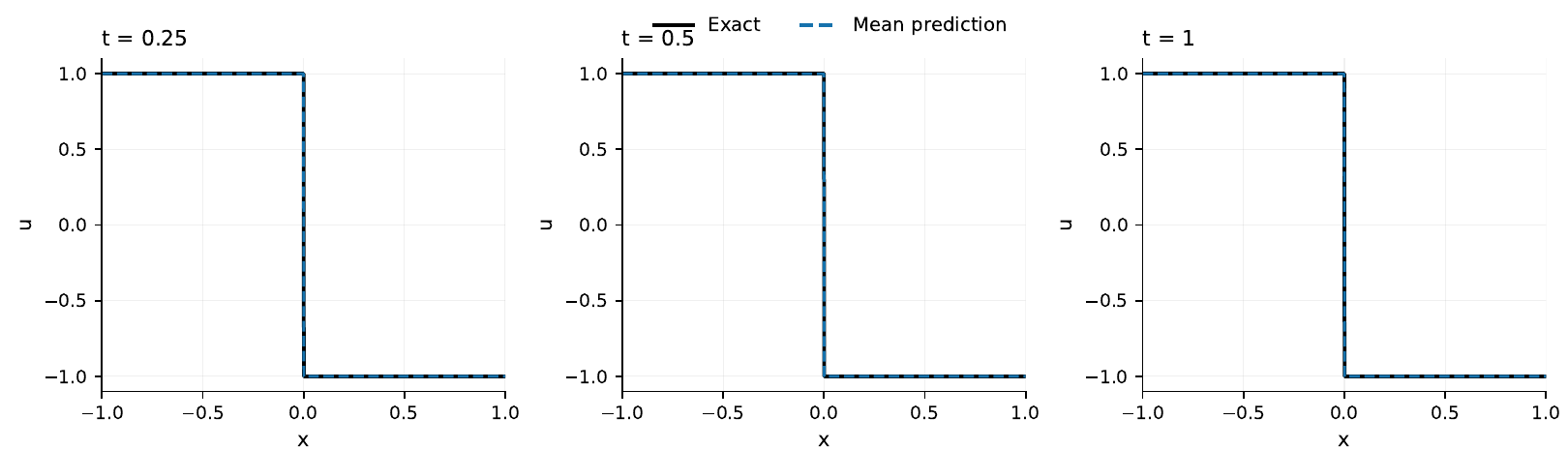}
\caption{Full-domain profiles of a stationary Burgers shock. Black solid lines denote the exact solution, and blue dashed lines denote the mean predictions of seeds 2, 3, 4, and 5 under the original configuration.}\label{fig:S1}
\end{figure}

\begin{figure}[!htbp]\centering
\includegraphics[width=\linewidth]{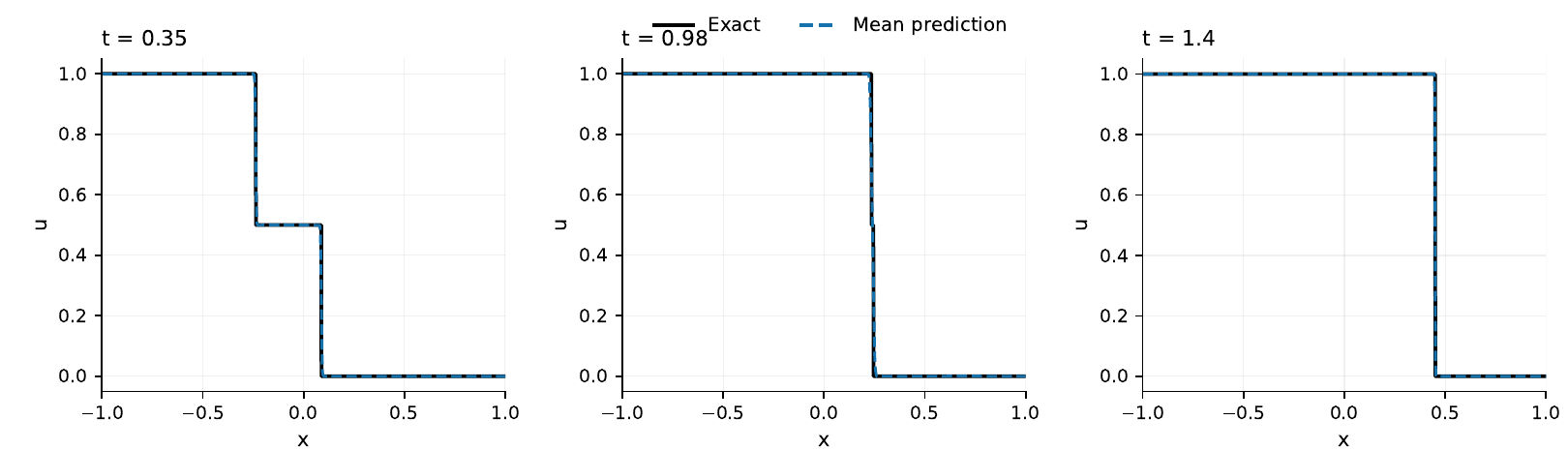}
\caption{Full-domain profiles before and after the merger of two Burgers shocks. Black solid lines denote the exact solution, and blue dashed lines denote the mean predictions of the four seeds with lower terminal errors under the original configuration.}\label{fig:S2}
\end{figure}

\begin{figure}[!htbp]\centering
\includegraphics[width=\linewidth]{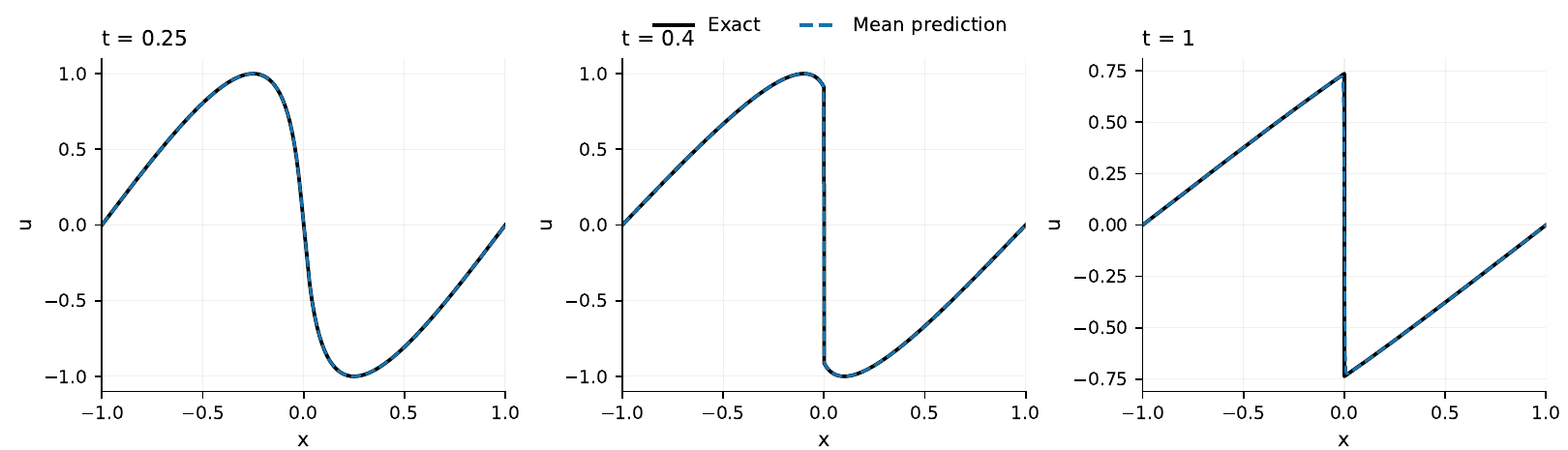}
\caption{Full-domain profiles of shock formation from sinusoidal initial data. Black solid lines denote the exact solution, and blue dashed lines denote the mean predictions of the four seeds with lower terminal errors under the time-slab restart scheme.}\label{fig:S3}
\end{figure}

\begin{figure}[!htbp]\centering
\includegraphics[width=\linewidth]{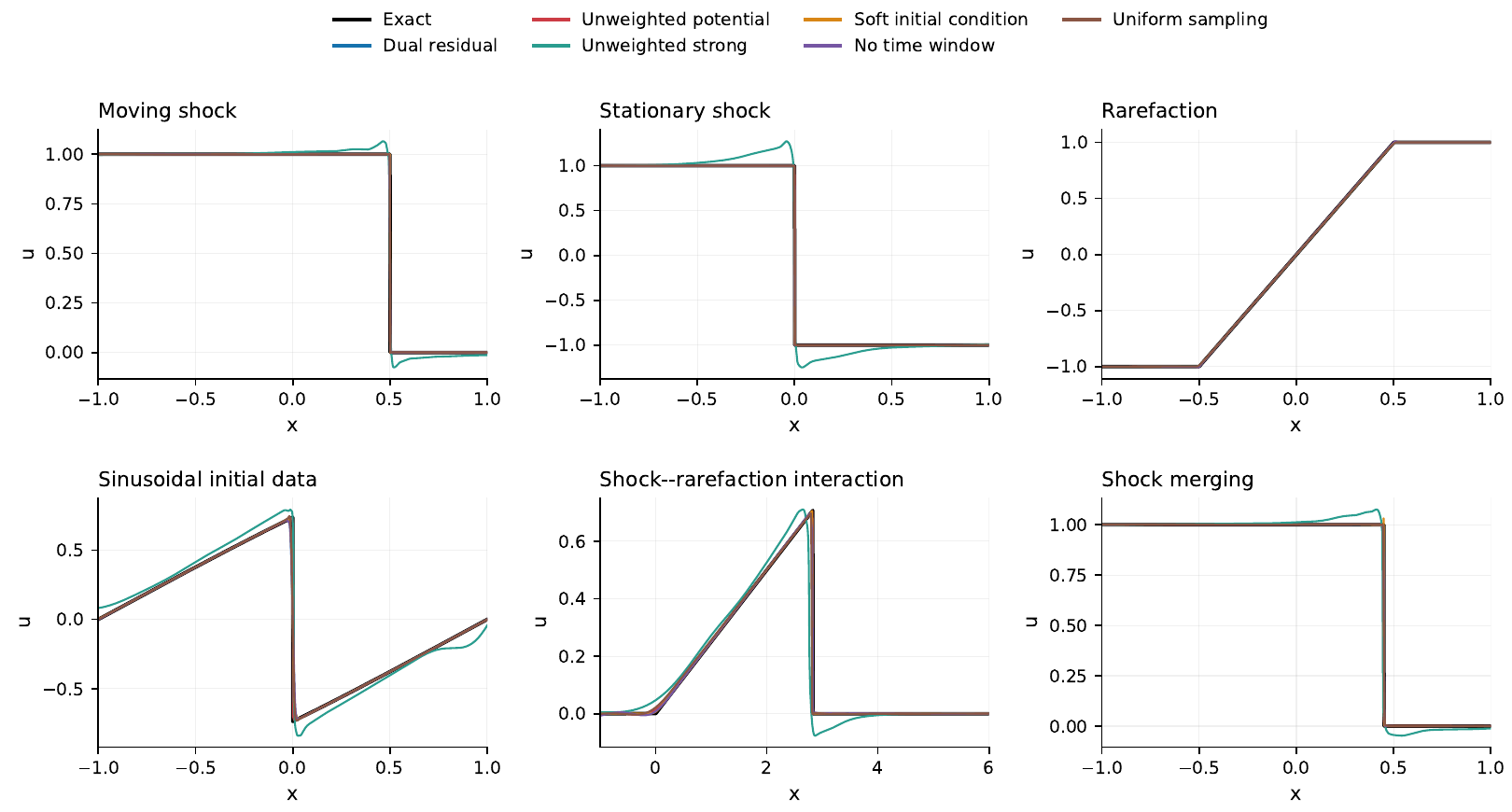}
\caption{Supplementary ablations for six Burgers initial conditions, all using seed~1. Black solid lines denote exact solutions; colored curves vary residual downweighting, the initial-data constraint, the time window, or the sampling rule. The rarefaction is shown at half-time and the other cases at the terminal time.}\label{fig:S4}
\end{figure}

\begin{figure}[!htbp]\centering
\includegraphics[width=\linewidth]{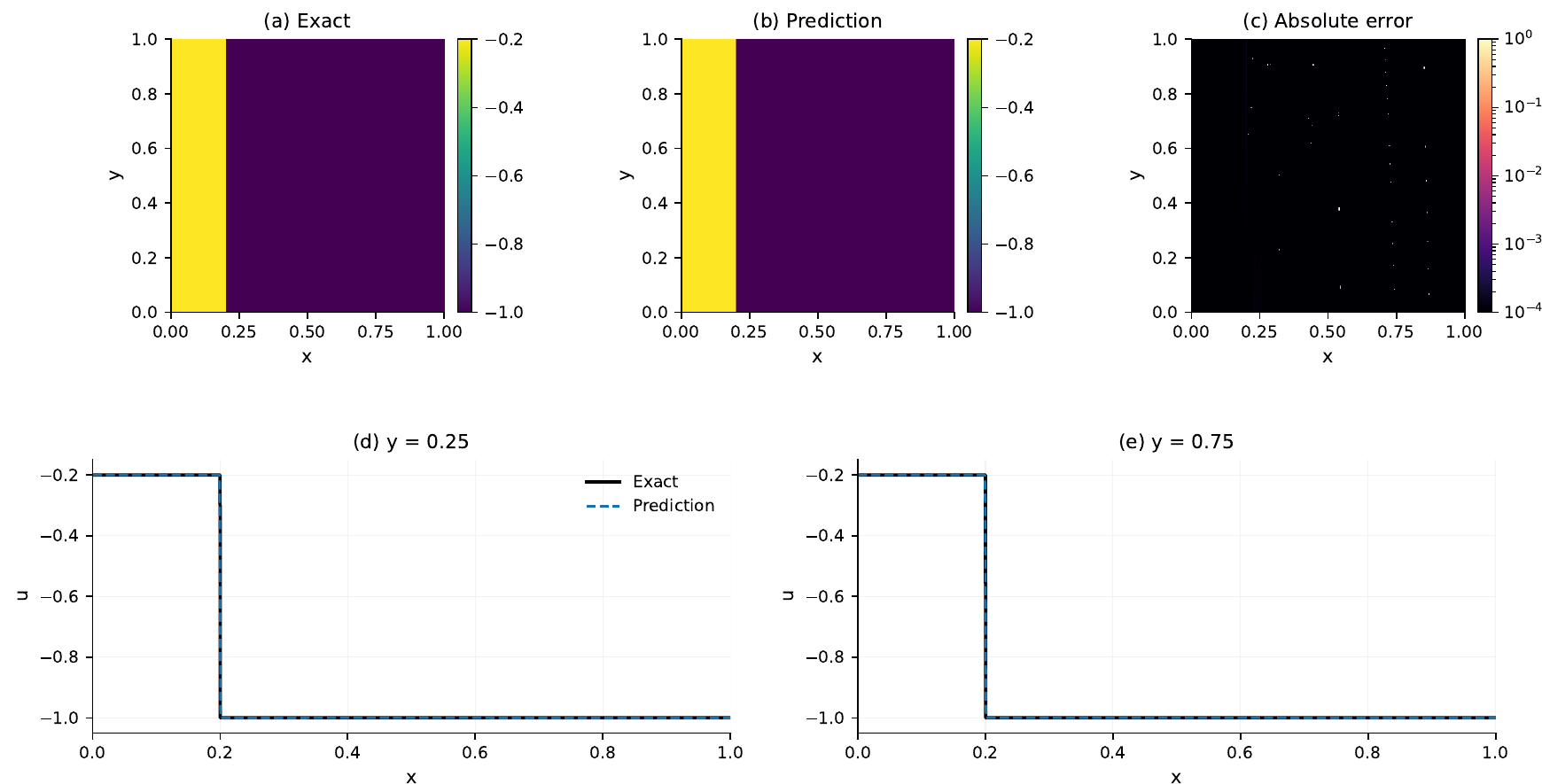}
\caption{Seed~1 results for the two-dimensional planar shock at $T=0.5$. The top row shows the exact field, prediction, and absolute error; the bottom row gives sections at $y=0.25$ and $y=0.75$. Black solid lines denote the exact solution and blue dashed lines the prediction. Table~4 reports refined-quadrature errors that resolve the transition layer.}\label{fig:S5}
\end{figure}

\begin{figure}[!htbp]\centering
\includegraphics[width=\linewidth]{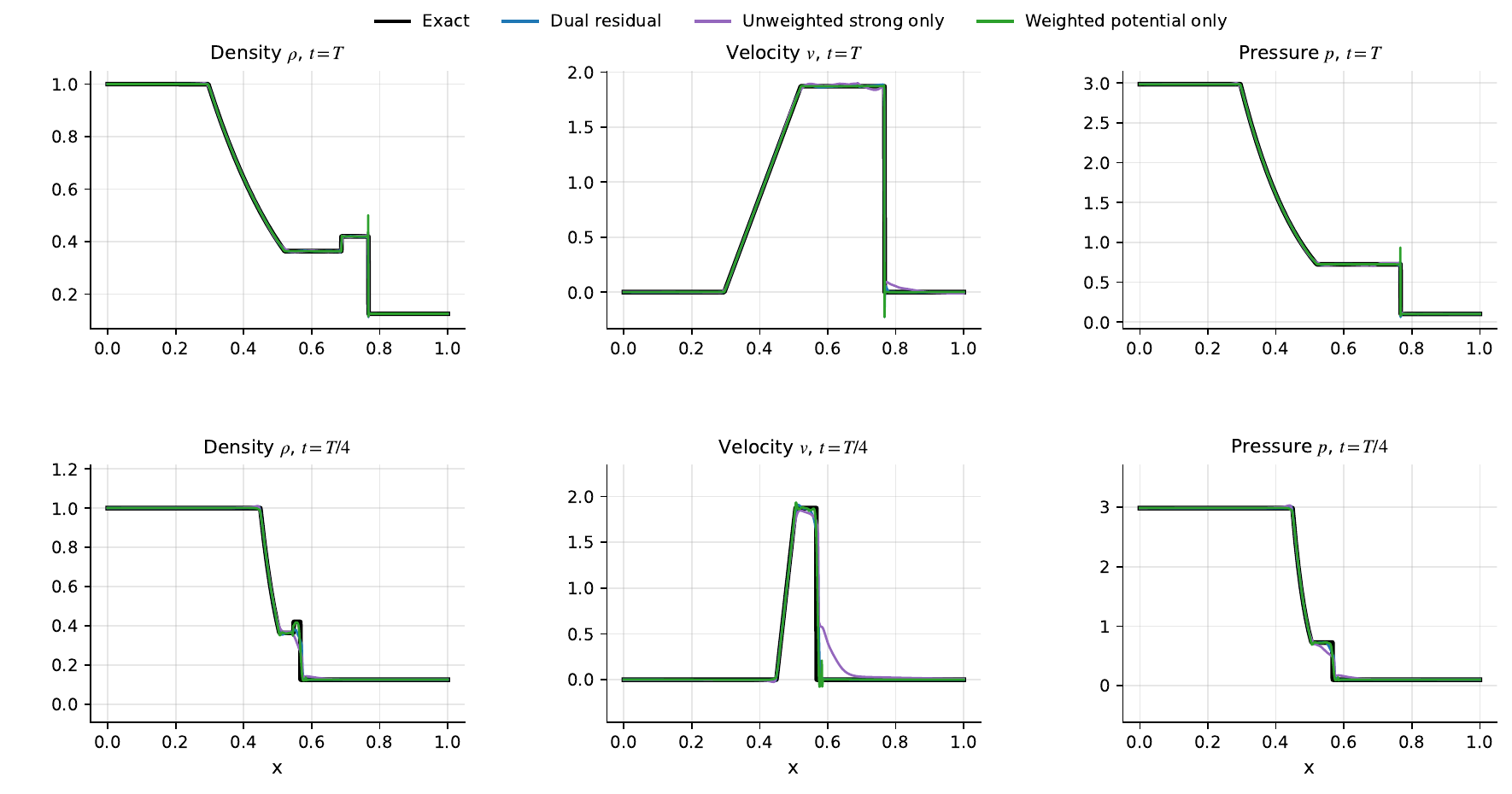}
\caption{Profiles from the independent low-pressure-ratio ablation, using 25 labels, seed~1, and left pressure approximately 3. The top row is terminal, and the bottom row is at the unlabeled time $T/4$; columns show density, velocity, and pressure. Black solid lines denote the exact solution. The comparison includes the complete dual-residual model, the unweighted strong-residual-only model, and the downweighted potential-residual-only model. At the unsupervised time, vertical limits extend the reference range by 25\%; full overshoot statistics are reported in Table~6.}\label{fig:S6}
\end{figure}

\begin{figure}[!htbp]\centering
\includegraphics[width=\linewidth]{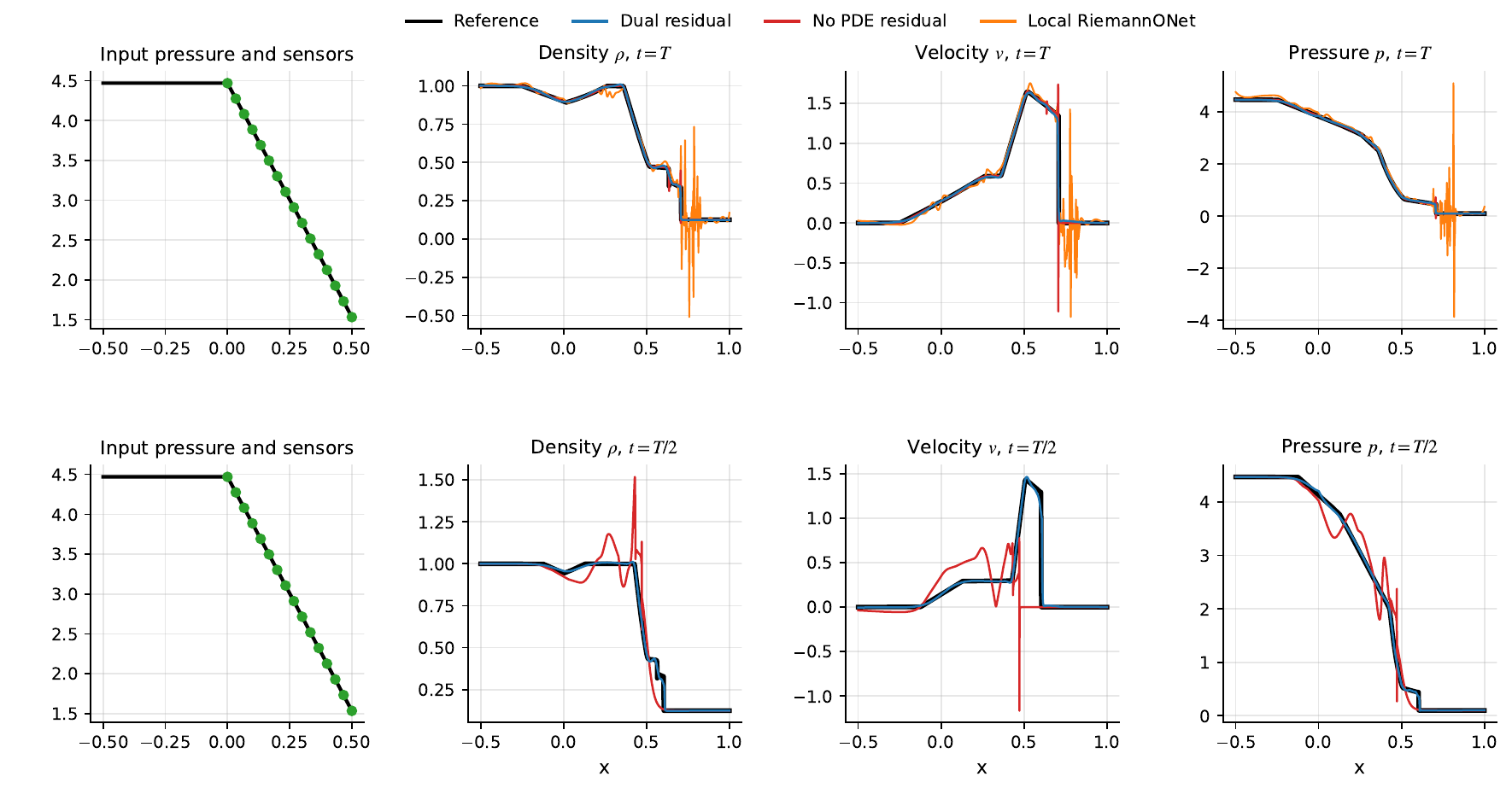}
\caption{Conditional predictions for a decreasing pressure function (test function~57). The legend is the same as in Fig.~13, with 25 labels and seed~2.}\label{fig:S7}
\end{figure}

\begin{figure}[!htbp]\centering
\includegraphics[width=\linewidth]{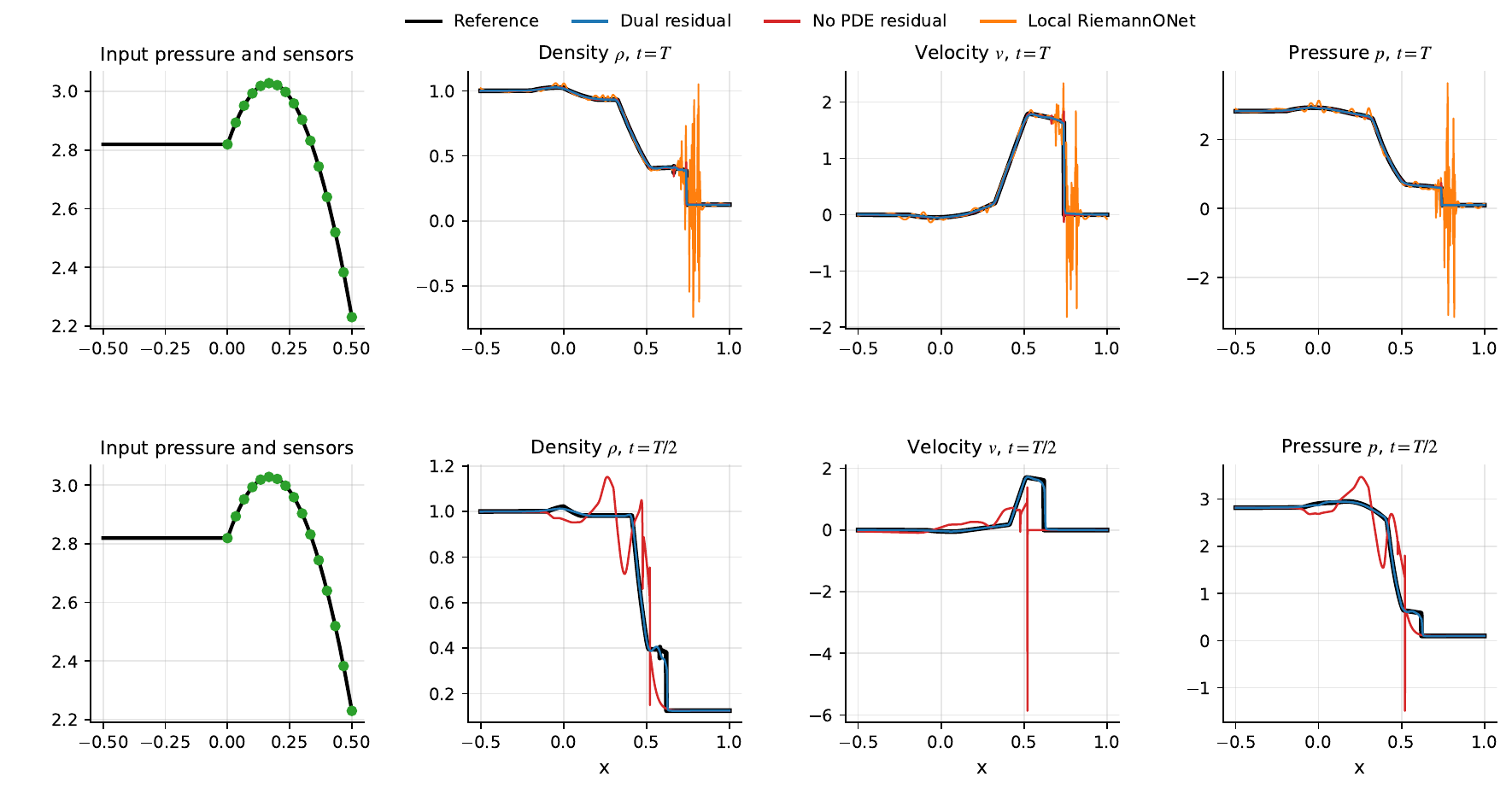}
\caption{Conditional predictions for a peaked pressure function (test function~35). The legend is the same as in Fig.~13, with 25 labels and seed~2.}\label{fig:S8}
\end{figure}

\begin{figure}[!htbp]\centering
\includegraphics[width=\linewidth]{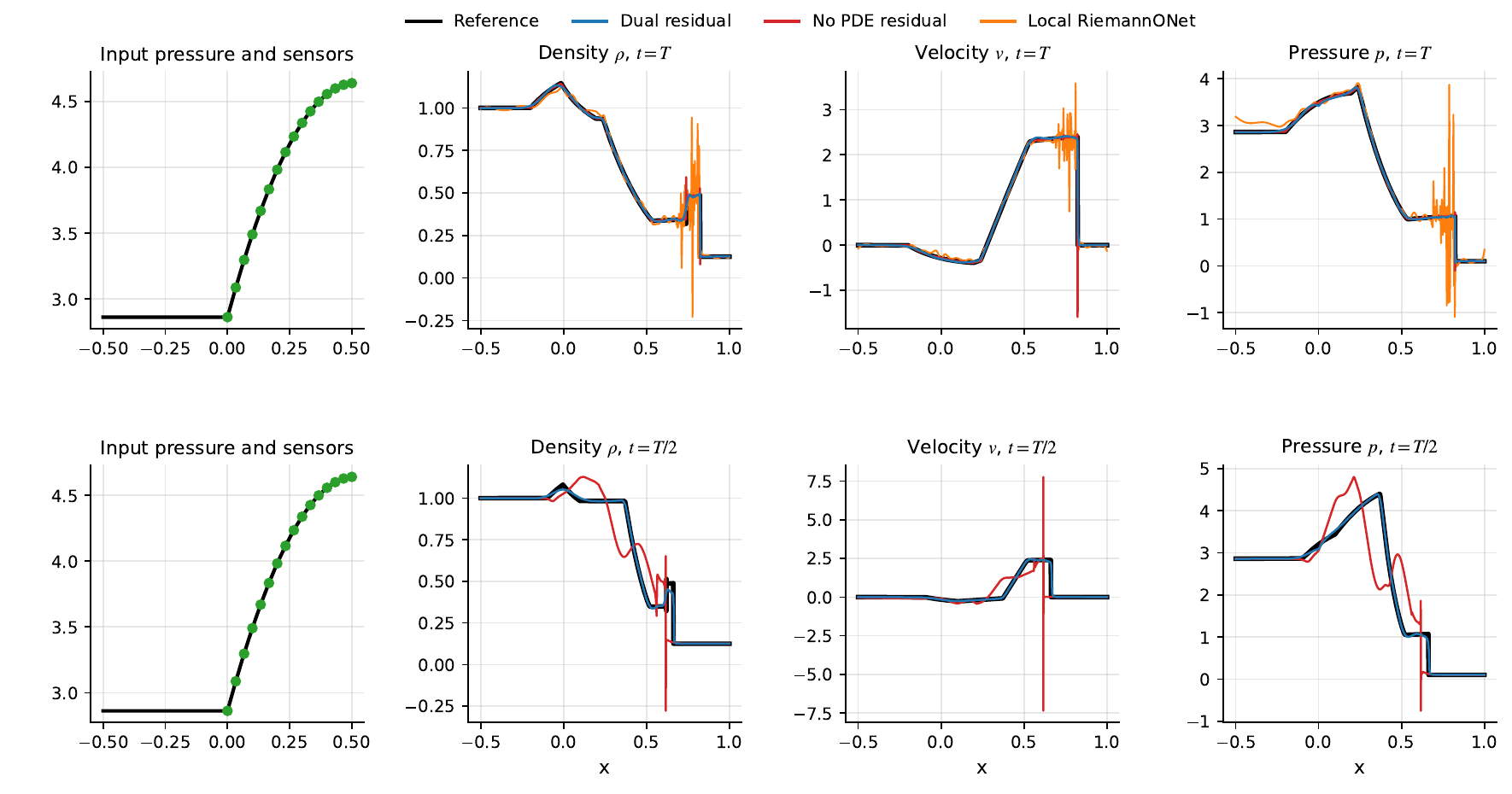}
\caption{Conditional predictions for another increasing pressure function (test function~24). The legend is the same as in Fig.~13, with 25 labels and seed~2.}\label{fig:S9}
\end{figure}

\begin{figure}[!htbp]\centering
\includegraphics[width=\linewidth]{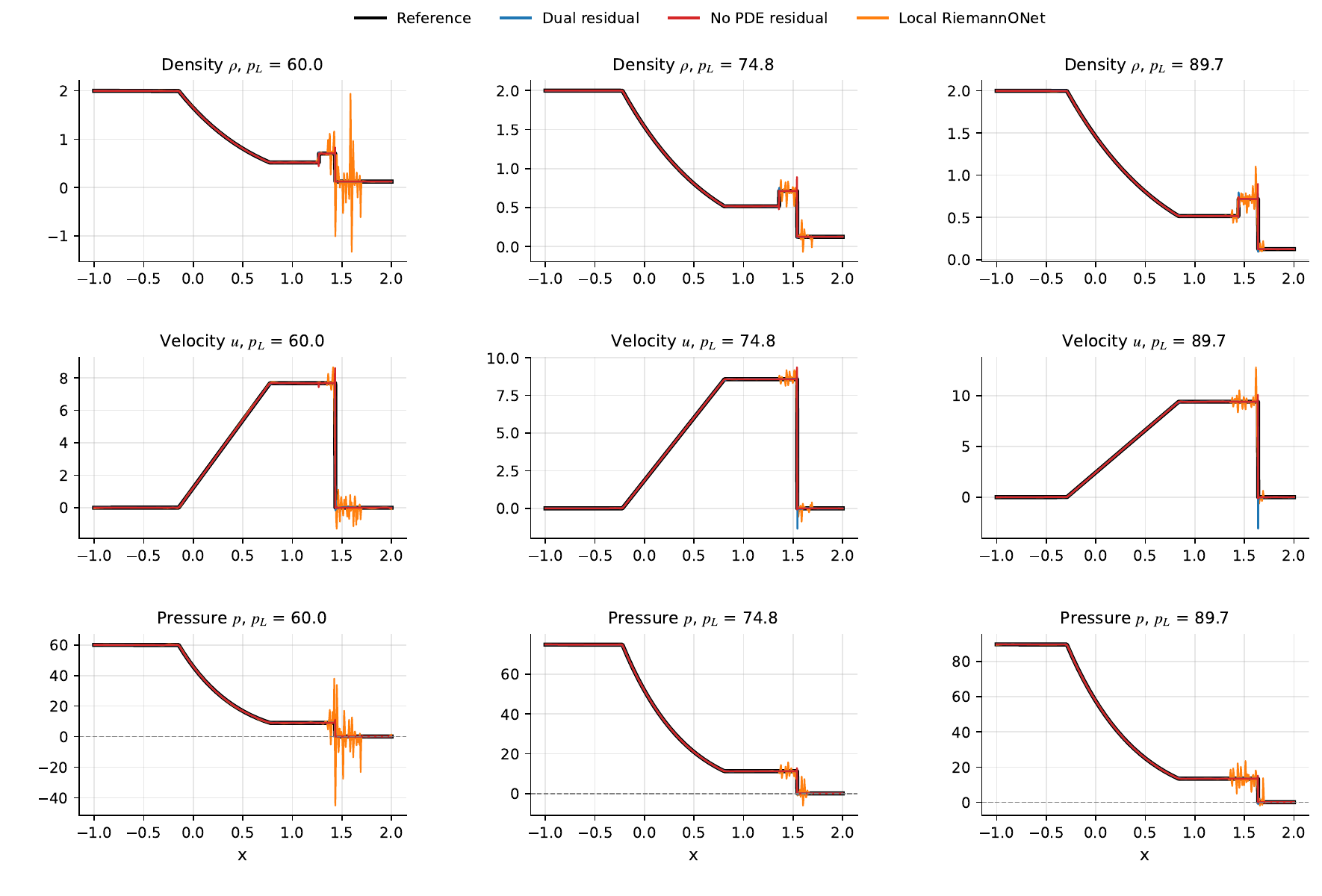}
\caption{Terminal profiles for the higher-pressure-ratio input family, using 25 labels and the common seed~3. Columns correspond to left pressures of approximately 60, 75, and 90; rows show density, velocity, and pressure. Black solid lines denote the exact solution. Vertical limits include all plotted values, and zero-pressure lines identify negative pressures.}\label{fig:S10}
\end{figure}

%% file: main.bbl
\begin{thebibliography}{99}

\bibitem[Bhattacharya et al.(2021)]{r33}
Bhattacharya, K.; Hosseini, B.; Kovachki, N. B.; Stuart, A. M. 2021. \href{https://doi.org/10.5802/smai-jcm.74}{Model reduction and neural networks for parametric PDEs.} \emph{SMAI J. Comput. Math. 7, 121--157.}

\bibitem[Chen and Chen(1995)]{r32}
Chen, T.; Chen, H. 1995. \href{https://doi.org/10.1109/72.392253}{Universal approximation to nonlinear operators by neural networks with arbitrary activation functions and its application to dynamical systems.} \emph{IEEE Trans. Neural Netw. 6, 911--917.}

\bibitem[Cuomo et al.(2022)]{r12}
Cuomo, S.; Di Cola, V. S.; Giampaolo, F.; Rozza, G.; Raissi, M.; Piccialli, F. 2022. \href{https://doi.org/10.1007/s10915-022-01939-z}{Scientific machine learning through physics-informed neural networks: Where we are and what's next.} \emph{J. Sci. Comput. 92, 88.}

\bibitem[De Ryck et al.(2024)]{r26}
De Ryck, T.; Mishra, S.; Molinaro, R. 2024. \href{https://doi.org/10.1137/22M1522504}{wPINNs: Weak physics informed neural networks for approximating entropy solutions of hyperbolic conservation laws.} \emph{SIAM J. Numer. Anal. 62, 811--841.}

\bibitem[Fresca et al.(2021)]{r43}
Fresca, S.; Dede', L.; Manzoni, A. 2021. \href{https://doi.org/10.1007/s10915-021-01462-7}{A comprehensive deep learning-based approach to reduced order modeling of nonlinear time-dependent parametrized PDEs.} \emph{J. Sci. Comput. 87, 61.}

\bibitem[Godunov(1959)]{r03}
Godunov, S. K. 1959. \href{https://www.mathnet.ru/eng/sm4873}{A difference method for numerical calculation of discontinuous solutions of the equations of hydrodynamics.} \emph{Mat. Sb. 47, 271--306.}

\bibitem[Guermond et al.(2011)]{r05}
Guermond, J.-L.; Pasquetti, R.; Popov, B. 2011. \href{https://doi.org/10.1016/j.jcp.2010.11.043}{Entropy viscosity method for nonlinear conservation laws.} \emph{J. Comput. Phys. 230, 4248--4267.}

\bibitem[Herde et al.(2024)]{r50}
Herde, M.; Raoni\'{c}, B.; Rohner, T.; K\"{a}ppeli, R.; Molinaro, R.; de B\'{e}zenac, E.; Mishra, S. 2024. \href{https://arxiv.org/abs/2405.19101}{Poseidon: Efficient foundation models for PDEs.} \emph{Advances in Neural Information Processing Systems 37.}

\bibitem[Jagtap et al.(2020)]{r23}
Jagtap, A. D.; Kharazmi, E.; Karniadakis, G. E. 2020. \href{https://doi.org/10.1016/j.cma.2020.113028}{Conservative physics-informed neural networks on discrete domains for conservation laws: Applications to forward and inverse problems.} \emph{Comput. Methods Appl. Mech. Engrg. 365, 113028.}

\bibitem[Ji et al.(2026)]{r47}
Ji, Y.; Chen, Y.; Xu, Z. 2026. \href{https://doi.org/10.1016/j.camwa.2026.09.017}{EGPT-PINN: Entropy-regularized generative pre-trained physics informed neural networks for parameterized nonlinear conservation laws.} \emph{Comput. Math. Appl. 222, 223--243.}

\bibitem[Jiang and Shu(1996)]{r04}
Jiang, G.-S.; Shu, C.-W. 1996. \href{https://doi.org/10.1006/jcph.1996.0130}{Efficient implementation of weighted ENO schemes.} \emph{J. Comput. Phys. 126, 202--228.}

\bibitem[Karniadakis et al.(2021)]{r11}
Karniadakis, G. E.; Kevrekidis, I. G.; Lu, L.; Perdikaris, P.; Wang, S.; Yang, L. 2021. \href{https://doi.org/10.1038/s42254-021-00314-5}{Physics-informed machine learning.} \emph{Nat. Rev. Phys. 3, 422--440.}

\bibitem[Kharazmi et al.(2021)]{r25}
Kharazmi, E.; Zhang, Z.; Karniadakis, G. E. 2021. \href{https://doi.org/10.1016/j.cma.2020.113547}{hp-VPINNs: Variational physics-informed neural networks with domain decomposition.} \emph{Comput. Methods Appl. Mech. Engrg. 374, 113547.}

\bibitem[Kovachki et al.(2021)]{r38}
Kovachki, N.; Lanthaler, S.; Mishra, S. 2021. \href{https://jmlr.org/papers/v22/21-0806.html}{On universal approximation and error bounds for Fourier neural operators.} \emph{J. Mach. Learn. Res. 22, 290, 1--76.}

\bibitem[Kovachki et al.(2023)]{r34}
Kovachki, N.; Li, Z.; Liu, B.; Azizzadenesheli, K.; Bhattacharya, K.; Stuart, A.; Anandkumar, A. 2023. \href{https://jmlr.org/papers/v24/21-1524.html}{Neural operator: Learning maps between function spaces with applications to PDEs.} \emph{J. Mach. Learn. Res. 24, 89, 1--97.}

\bibitem[Krishnapriyan et al.(2021)]{r17}
Krishnapriyan, A.; Gholami, A.; Zhe, S.; Kirby, R.; Mahoney, M. W. 2021. \href{https://arxiv.org/abs/2109.01050}{Characterizing possible failure modes in physics-informed neural networks.} \emph{Advances in Neural Information Processing Systems 34.}

\bibitem[Kru\v{z}kov(1970)]{r02}
Kru\v{z}kov, S. N. 1970. \href{https://www.mathnet.ru/eng/sm3372}{First order quasilinear equations in several independent variables.} \emph{Math. USSR-Sb. 10, 217--243.}

\bibitem[Lagaris et al.(1998)]{r08}
Lagaris, I. E.; Likas, A.; Fotiadis, D. I. 1998. \href{https://doi.org/10.1109/72.712178}{Artificial neural networks for solving ordinary and partial differential equations.} \emph{IEEE Trans. Neural Netw. 9, 987--1000.}

\bibitem[Lanthaler et al.(2022)]{r37}
Lanthaler, S.; Mishra, S.; Karniadakis, G. E. 2022. \href{https://doi.org/10.1093/imatrm/tnac001}{Error estimates for DeepONets: A deep learning framework in infinite dimensions.} \emph{Trans. Math. Appl. 6, tnac001.}

\bibitem[Lanthaler et al.(2023)]{r45}
Lanthaler, S.; Molinaro, R.; Hadorn, P.; Mishra, S. 2023. \href{https://openreview.net/forum?id=CrfhZAsJDsZ}{Nonlinear reconstruction for operator learning of PDEs with discontinuities.} \emph{International Conference on Learning Representations.}

\bibitem[Lax(1954)]{r01}
Lax, P. D. 1954. \href{https://doi.org/10.1002/cpa.3160070112}{Weak solutions of nonlinear hyperbolic equations and their numerical computation.} \emph{Commun. Pure Appl. Math. 7, 159--193.}

\bibitem[Li et al.(2021)]{r36}
Li, Z.; Kovachki, N.; Azizzadenesheli, K.; Liu, B.; Bhattacharya, K.; Stuart, A.; Anandkumar, A. 2021. \href{https://openreview.net/forum?id=c8P9NQVtmnO}{Fourier neural operator for parametric partial differential equations.} \emph{International Conference on Learning Representations.}

\bibitem[Li et al.(2023)]{r42}
Li, Z.; Zheng, H.; Kovachki, N.; Jin, D.; Chen, H.; Liu, B.; Azizzadenesheli, K.; Anandkumar, A. 2023. \href{https://arxiv.org/abs/2111.03794v4}{Physics-informed neural operator for learning partial differential equations.} \emph{arXiv:2111.03794v4.}

\bibitem[Liu et al.(2024a)]{r22}
Liu, L.; Liu, S.; Xie, H.; Xiong, F.; Yu, T.; Xiao, M.; Liu, L.; Yong, H. 2024a. \href{https://doi.org/10.1007/s10915-023-02412-1}{Discontinuity computing using physics-informed neural networks.} \emph{J. Sci. Comput. 98, 22.}

\bibitem[Liu et al.(2026)]{r28}
Liu, M.; Peng, G.; Xiong, F.; Zou, Q. 2026. \href{https://doi.org/10.2139/ssrn.6212057}{HyDFVM: A hybrid finite-volume learning for hyperbolic equations with discontinuous solutions.} \emph{SSRN preprint 6212057.}

\bibitem[Liu et al.(2024b)]{r48}
Liu, N.; Fan, Y.; Zeng, X.; Kl\"{o}wer, M.; Zhang, L.; Yu, Y. 2024b. \href{https://proceedings.mlr.press/v235/liu24p.html}{Harnessing the power of neural operators with automatically encoded conservation laws.} \emph{Proc. Mach. Learn. Res. 235, 30965--30997.}

\bibitem[Lu et al.(2021a)]{r35}
Lu, L.; Jin, P.; Pang, G.; Zhang, Z.; Karniadakis, G. E. 2021a. \href{https://doi.org/10.1038/s42256-021-00302-5}{Learning nonlinear operators via DeepONet based on the universal approximation theorem of operators.} \emph{Nat. Mach. Intell. 3, 218--229.}

\bibitem[Lu et al.(2021b)]{r13}
Lu, L.; Meng, X.; Mao, Z.; Karniadakis, G. E. 2021b. \href{https://doi.org/10.1137/19M1274067}{DeepXDE: A deep learning library for solving differential equations.} \emph{SIAM Rev. 63, 208--228.}

\bibitem[Mao et al.(2020)]{r14}
Mao, Z.; Jagtap, A. D.; Karniadakis, G. E. 2020. \href{https://doi.org/10.1016/j.cma.2019.112789}{Physics-informed neural networks for high-speed flows.} \emph{Comput. Methods Appl. Mech. Engrg. 360, 112789.}

\bibitem[McGreivy and Hakim(2024)]{r51}
McGreivy, N.; Hakim, A. 2024. \href{https://doi.org/10.1038/s42256-024-00897-5}{Weak baselines and reporting biases lead to overoptimism in machine learning for fluid-related partial differential equations.} \emph{Nat. Mach. Intell. 6, 1256--1269.}

\bibitem[Oubarka et al.(2026)]{r27}
Oubarka, I.; Kissami, I.; Boubekeur, M.; Benkhaldoun, F.; Madrane, A.; Saadi, Z. 2026. \href{https://arxiv.org/abs/2603.24819v2}{Weak and entropy physics-informed neural networks for conservation laws.} \emph{arXiv:2603.24819v2.}

\bibitem[Patel et al.(2022)]{r24}
Patel, R. G.; Manickam, I.; Trask, N. A.; Wood, M. A.; Lee, M.; Tomas, I.; Cyr, E. C. 2022. \href{https://doi.org/10.1016/j.jcp.2021.110754}{Thermodynamically consistent physics-informed neural networks for hyperbolic systems.} \emph{J. Comput. Phys. 449, 110754.}

\bibitem[Peyvan et al.(2024)]{r46}
Peyvan, A.; Oommen, V.; Jagtap, A. D.; Karniadakis, G. E. 2024. \href{https://doi.org/10.1016/j.cma.2024.116996}{RiemannONets: Interpretable neural operators for Riemann problems.} \emph{Comput. Methods Appl. Mech. Engrg. 426, 116996.}

\bibitem[Raissi et al.(2019)]{r10}
Raissi, M.; Perdikaris, P.; Karniadakis, G. E. 2019. \href{https://doi.org/10.1016/j.jcp.2018.10.045}{Physics-informed neural networks: A deep learning framework for solving forward and inverse problems involving nonlinear partial differential equations.} \emph{J. Comput. Phys. 378, 686--707.}

\bibitem[Rajvanshi and Ketcheson(2024)]{r29}
Rajvanshi, M. P.; Ketcheson, D. I. 2024. \href{https://openreview.net/forum?id=Uuu6HWe6dF}{Integral PINNs for hyperbolic conservation laws.} \emph{ICLR Workshop on AI4Differential Equations in Science.}

\bibitem[Raoni\'{c} et al.(2023)]{r40}
Raoni\'{c}, B.; Molinaro, R.; De Ryck, T.; Rohner, T.; Bartolucci, F.; Alaifari, R.; Mishra, S.; de B\'{e}zenac, E. 2023. \href{https://arxiv.org/abs/2302.01178}{Convolutional neural operators for robust and accurate learning of PDEs.} \emph{Advances in Neural Information Processing Systems 36.}

\bibitem[Rathore et al.(2024)]{r18}
Rathore, P.; Lei, W.; Frangella, Z.; Lu, L.; Udell, M. 2024. \href{https://proceedings.mlr.press/v235/rathore24a.html}{Challenges in training PINNs: A loss landscape perspective.} \emph{Proc. Mach. Learn. Res. 235, 42159--42191.}

\bibitem[Seidman et al.(2022)]{r44}
Seidman, J. H.; Kissas, G.; Perdikaris, P.; Pappas, G. J. 2022. \href{https://arxiv.org/abs/2206.03551}{NOMAD: Nonlinear manifold decoders for operator learning.} \emph{Advances in Neural Information Processing Systems 35.}

\bibitem[Sirignano and Spiliopoulos(2018)]{r09}
Sirignano, J.; Spiliopoulos, K. 2018. \href{https://doi.org/10.1016/j.jcp.2018.08.029}{DGM: A deep learning algorithm for solving partial differential equations.} \emph{J. Comput. Phys. 375, 1339--1364.}

\bibitem[Sod(1978)]{r06}
Sod, G. A. 1978. \href{https://doi.org/10.1016/0021-9991(78)90023-2}{A survey of several finite difference methods for systems of nonlinear hyperbolic conservation laws.} \emph{J. Comput. Phys. 27, 1--31.}

\bibitem[Takamoto et al.(2022)]{r49}
Takamoto, M.; Praditia, T.; Leiteritz, R.; MacKinlay, D.; Alesiani, F.; Pfl\"{u}ger, D.; Niepert, M. 2022. \href{https://arxiv.org/abs/2210.07182}{PDEBench: An extensive benchmark for scientific machine learning.} \emph{Advances in Neural Information Processing Systems 35.}

\bibitem[Tran et al.(2023)]{r39}
Tran, A.; Mathews, A.; Xie, L.; Ong, C. S. 2023. \href{https://arxiv.org/abs/2111.13802}{Factorized Fourier neural operators.} \emph{International Conference on Learning Representations.}

\bibitem[Wang et al.(2023)]{r19}
Wang, S.; Sankaran, S.; Wang, H.; Perdikaris, P. 2023. \href{https://arxiv.org/abs/2308.08468}{An expert's guide to training physics-informed neural networks.} \emph{arXiv:2308.08468.}

\bibitem[Wang et al.(2021a)]{r15}
Wang, S.; Teng, Y.; Perdikaris, P. 2021a. \href{https://doi.org/10.1137/20M1318043}{Understanding and mitigating gradient flow pathologies in physics-informed neural networks.} \emph{SIAM J. Sci. Comput. 43, A3055--A3081.}

\bibitem[Wang et al.(2021b)]{r41}
Wang, S.; Wang, H.; Perdikaris, P. 2021b. \href{https://doi.org/10.1126/sciadv.abi8605}{Learning the solution operator of parametric partial differential equations with physics-informed DeepONets.} \emph{Sci. Adv. 7, eabi8605.}

\bibitem[Wang et al.(2022)]{r16}
Wang, S.; Yu, X.; Perdikaris, P. 2022. \href{https://doi.org/10.1016/j.jcp.2021.110768}{When and why PINNs fail to train: A neural tangent kernel perspective.} \emph{J. Comput. Phys. 449, 110768.}

\bibitem[Wang and Yang(2026)]{r30}
Wang, Y.; Yang, S. 2026. \href{https://arxiv.org/abs/2411.11276v2}{Coupled integral PINN for discontinuity.} \emph{arXiv:2411.11276v2.}

\bibitem[Wu et al.(2023)]{r20}
Wu, C.; Zhu, M.; Tan, Q.; Kartha, Y.; Lu, L. 2023. \href{https://doi.org/10.1016/j.cma.2022.115671}{A comprehensive study of non-adaptive and residual-based adaptive sampling for physics-informed neural networks.} \emph{Comput. Methods Appl. Mech. Engrg. 403, 115671.}

\bibitem[Xiong et al.(2022)]{r21}
Xiong, F.; Liu, L.; Liu, S.; Wang, H.; Yong, H. 2022. \href{https://doi.org/10.36227/techrxiv.20099957.v1}{Gradient-weighted physics-informed neural networks for one-dimensional Euler equation.} \emph{TechRxiv preprint.}

\bibitem[Yu et al.(2022)]{r31}
Yu, J.; Lu, L.; Meng, X.; Karniadakis, G. E. 2022. \href{https://doi.org/10.1016/j.cma.2022.114823}{Gradient-enhanced physics-informed neural networks for forward and inverse PDE problems.} \emph{Comput. Methods Appl. Mech. Engrg. 393, 114823.}

\bibitem[Zhang and Shu(2010)]{r07}
Zhang, X.; Shu, C.-W. 2010. \href{https://doi.org/10.1016/j.jcp.2010.08.016}{On positivity-preserving high order discontinuous Galerkin schemes for compressible Euler equations on rectangular meshes.} \emph{J. Comput. Phys. 229, 8918--8934.}

\end{thebibliography}
